\documentclass[lettersize,journal]{IEEEtran}
\usepackage{amsmath,amsfonts}
\usepackage{algorithmic}
\usepackage{algorithm}
\usepackage{array}
\usepackage[caption=false,font=normalsize,labelfont=sf,textfont=sf]{subfig}
\usepackage{textcomp}
\usepackage{stfloats}
\usepackage{url}
\usepackage{verbatim}
\usepackage{graphicx}
\usepackage{cite}
\begin{document}

\title{A CNN-Based Technique to Assist Layout-to-Generator Conversion for Analog Circuits}

\author{Sungyu Jeong,~\IEEEmembership{Graduate Student Member,~IEEE,}
        Minsu Kim,~\IEEEmembership{Graduate Student Member,~IEEE,}

        and Byungsub Kim,~\IEEEmembership{Senior Member,~IEEE}

        % <-this % stops a space
\thanks{This work was supported in part by Institute of Information and Communications Technology Planning and Evaluation grant funded by the Korea Government (MSIT) (No. 2022-0-01171);
in part by BK21 FOUR Project of NRF for the Department of Electrical Engineering, POSTECH;
in part by Next-generation Intelligence semiconductor R\&D Program through the National Research Foundation of Korea(NRF) funded by Korea goverment(MSIT) (RS-2023-00258227).
}
\thanks{Sungyu Jeong and Minsu Kim are with the Department of Electrical Engineering, Pohang University of Science and Technology, Pohang-si 37673, South Korea.}
\thanks{Byungsub Kim is with the Department of Electrical Engineering, the Department of Convergence IT Engineering, and the Department of Semiconductor Engineering, Pohang University of Science and Technology, Pohang-si 37673, South Korea, and also with the Institute for Convergence Research and Education in Advanced Technology, Yonsei University, Seoul 03722, South Korea (e-mail: byungsub@postech.ac.kr).}
}

% \markboth{}

\maketitle


\begin{abstract}
We propose a technique to assist in converting a reference layout of an analog circuit into the procedural layout generator by efficiently reusing available generators for sub-cell creation. The proposed convolutional neural network (CNN) model automatically detects sub-cells that can be generated by available generator scripts in the library, and suggests using them in the hierarchically correct places of the generator software. In experiments, the CNN model examined sub-cells of a high-speed wireline receiver that has a total of 4,885 sub-cell instances including different 145 sub-cell designs. The CNN model classified the sub-cell instances into 51 generatable and one not-generatable classes. One not-generatable class indicates that no available generator can generate the classified sub-cell. The CNN model achieved 99.3\% precision in examining the 145 different sub-cell designs. The CNN model greatly reduced the examination time to 18 seconds from 88 minutes required in manual examination. Also, the proposed CNN model could correctly classify unfamiliar sub-cells that are very different from the training dataset.
\end{abstract}


% \newpage
\section{Introduction}
\IEEEPARstart{D}{esigning} layouts of analog circuits is a time-consuming and costly process. Because the performances of analog circuits can be seriously degraded by poor physical designs, designers must spend many hours carefully drawing analog layouts considering the impacts of physical designs on analog performances. For example, the accuracy of an analog-to-digital converter (ADC) can be seriously degraded if capacitor layouts are poorly matched. Similarly, the speed of a serial link might be limited by the skew caused by poor high-speed routings and parasitic components. Therefore, designing analog layouts demands a lot of design time, the designer’s attention to detail, and their expertise in analog circuits.

\begin{figure}[b]
\centering
\includegraphics[width=\linewidth]{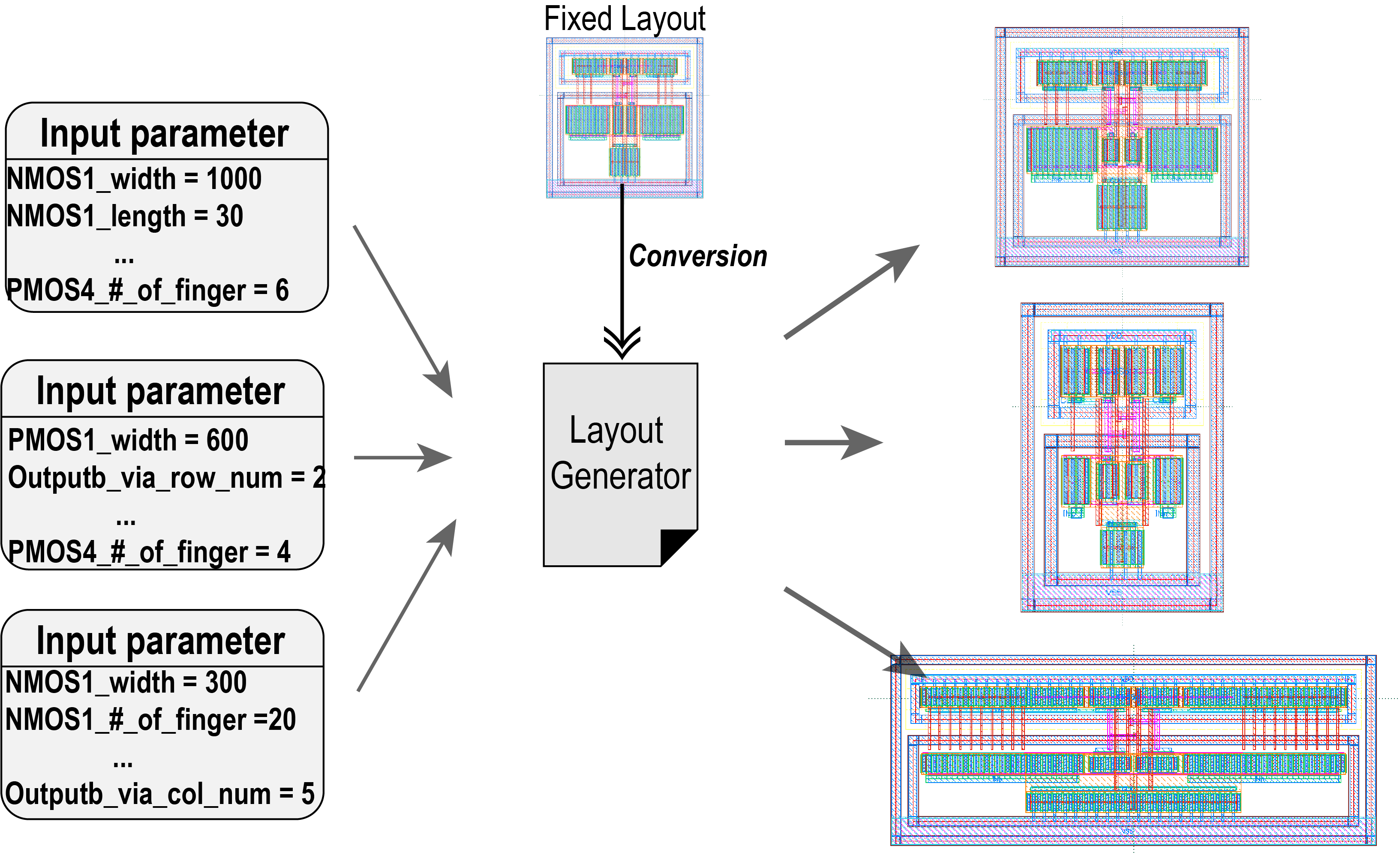}
\caption{The concept of layout-to-generator conversion, and example generation of sense-amplifiers using the procedural generator converted from the reference sense amplifier layout. For different input parameters, many different layout geometries such as the width/length/count of fingers and the number of vias can be modified.}
\label{fig_1_concept}
\end{figure}

To reduce analog layout development time and cost, researchers proposed “procedural layout generators” \cite{BAG, LAYGO, HAN}. A procedural layout generator receives input parameters from a designer and then automatically generates a layout instance (Fig. 1). This is especially useful when a design needs multiple layouts with different device sizes, or when frequent modification/revision is required. Also, it can be used when porting a layout design to different technology nodes by modifying the physical layers and adjusting geometric dimensions. Therefore, once it is developed, a procedural layout generator can significantly reduce the time and expense of analog layout development.

However, programming a procedural layout generator also takes a lot of time, expense, and skilled personnel. A generator script must be carefully programmed, considering analog performance and design rules. Therefore, this process also requires a designer with both knowledge of layouts and programming skills. Such highly talented engineers are expensive.

\begin{figure}[t]
\centering
\includegraphics[width=\linewidth]{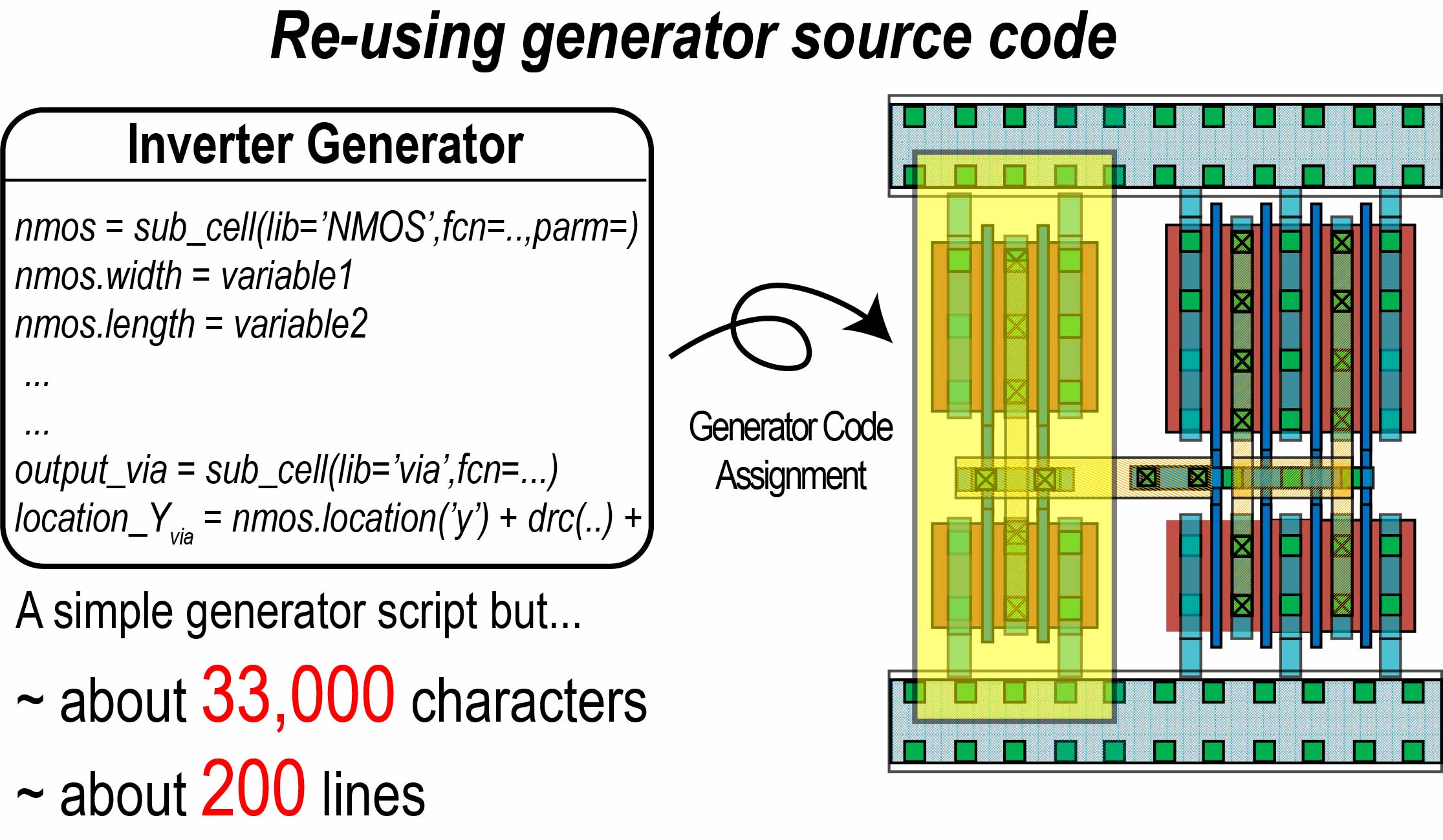}
\caption{A reusing example of an inverter generator script during the development of the procedural generator of an inverter chain layout.}
\label{fig2_reuse}
\end{figure}

\begin{figure}[t]
\centering
\includegraphics[width=\linewidth]{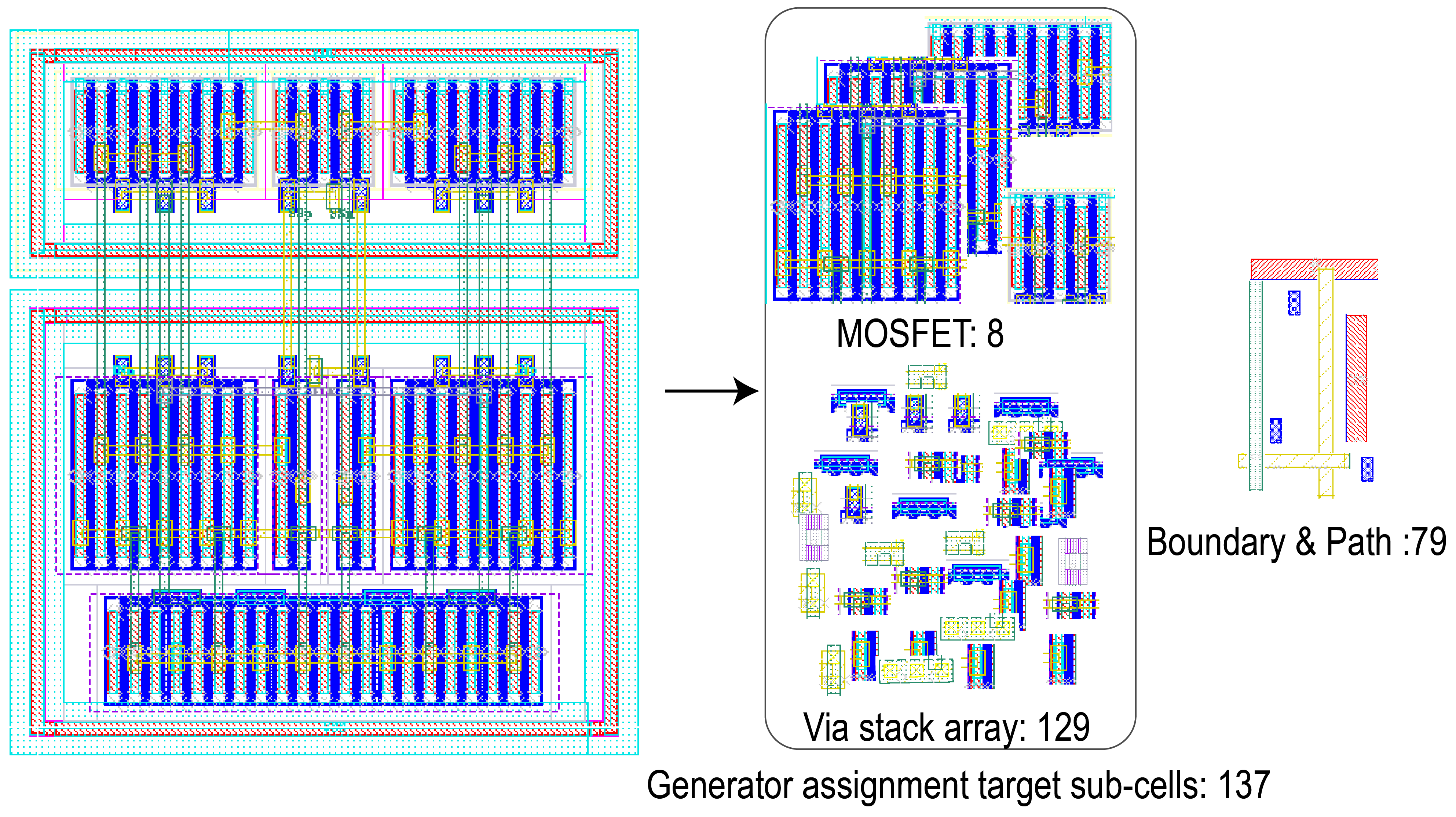}
\caption{An example of sub-cell layouts that can be generated with existing generator scripts. The example sense amplifier layout consists of 8 MOSFETs, 129 via stacks, and 79 boundary/path elements. The correct sub-cell generators must be assigned to 8 MOSFETs and 129 via stacks to develop the generator of the sense-amplifier.}
\label{fig3_subcell}
\end{figure}

This work aims at layout-to-generator conversion to reduce the development time and cost for the procedural layout generator. In the application, a designer can load an existing reference layout to reutilize the carefully planned layout styles in programming the procedural generator \cite{SUN}. By referencing the loaded layout, the designer can easily program the generator in a much shorter time. In addition, by referencing a silicon-proved layout, a novice designer can develop a high-quality layout generator without much knowledge about analog layout design \cite{SUN}.

Layout-to-generator conversion is done by appropriately expressing the geometries of the reference layout with parameters in the program code. If the reference layout has sub-cells that can be generated by available layout generators, then the amount of work can be greatly reduced by reusing the program code of the available generators. A large amount of its source code can be automatically loaded in and referred by the program under development in order to minimize the tedious typing of complex code. Fig. 2 shows an example of reusing an inverter generator script in developing an inverter chain generator. About 33,000 characters of the source code of the inverter generator can be reused in the source code of the inverter chain by referring to it.

However, a large layout may have many sub-cells, and examining the match between many sub-cells and many available generators also requires a lot of work. For example, Fig. 3 shows a layout of a simple sense amplifier and its layout elements. It has 137 sub-cells (8 MOSFETs and 129 via stacks), and 79 boundary/path elements. A larger layout may have many more sub-cells including more complex ones. In such a case, examining and assigning correct generators to sub-cells would be time-consuming.

We propose a CNN-based assistance technique to reduce the time-consuming labor of sub-cell generator assignment during layout-to-generator conversion. When a reference layout is loaded, our tool traces through the design hierarchy of the reference layout, and the CNN model examines each sub-cell, and suggests the generator script that can generate the sub-cell. Upon the designer’s approval, the matched generator scripts are automatically assigned to the right sub-cells. In experiments, the CNN model achieved 99.3\% precision in classifying the sub-cell designs of the reference layout, and properly suggested the generator scripts that can generate the correctly classified ones. Compared with the manual examination, the working time was reduced by the CNN model from 88 minutes to 18 seconds in classifying 4,885 sub-cell instances into 52 classes. The time improvement would be much greater in a practical application where there are much more number of sub-cells and available generators.

\section{Background}
\subsection{Procedural Layout Geneartor}

\begin{figure}[b]
        \centering
        \includegraphics[width=\linewidth]{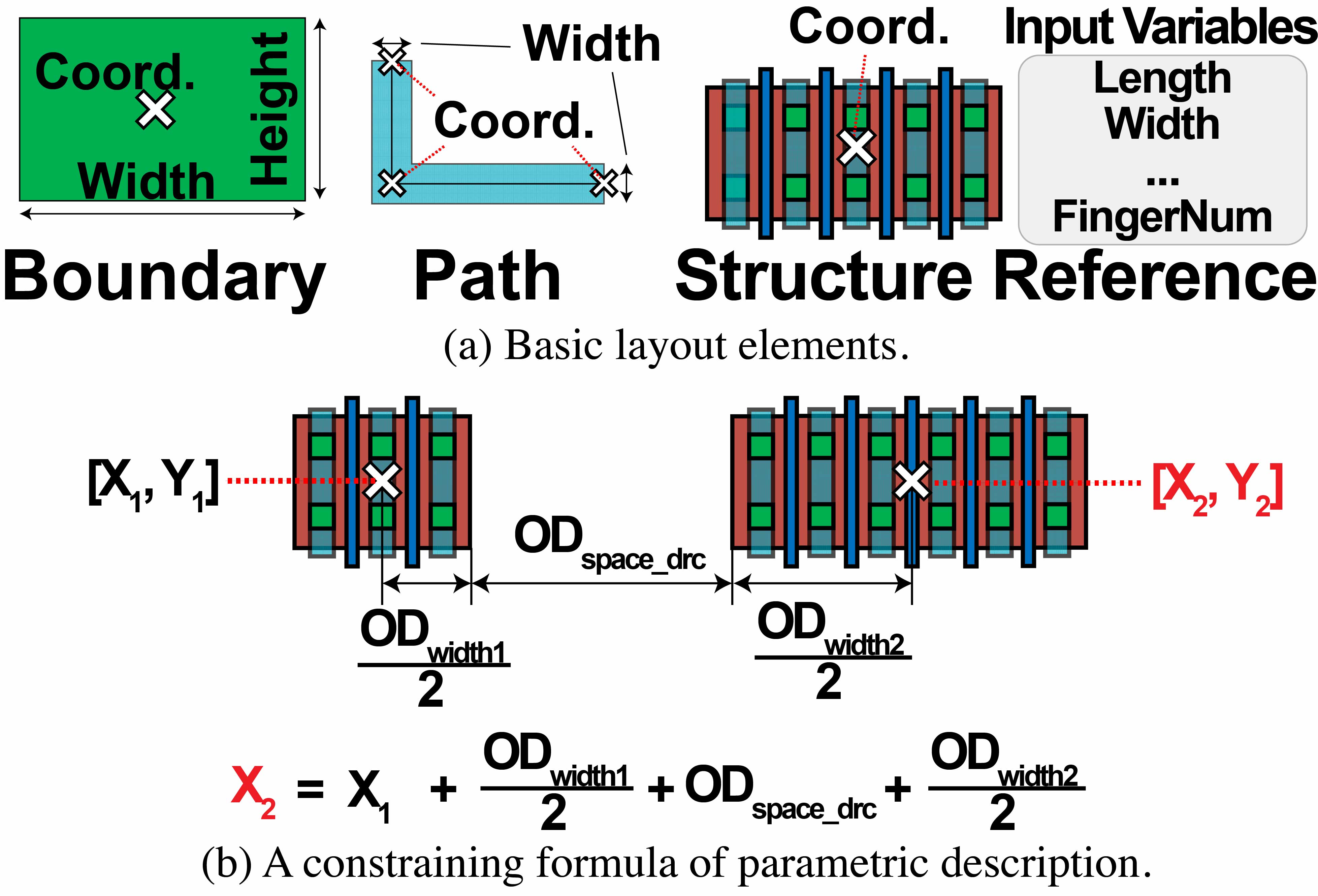}
        \caption{Basic Layout elements and parametric description formula. Reprinted from [4].}
\end{figure}
        
Procedural layout generators can automatically create layout designs for specific circuit types by receiving design parameters as input. Instead of manually creating layouts for each design iteration, designers can quickly generate multiple layout variants by adjusting input parameters. These generator scripts are particularly valuable for analog and mixed-signal circuit blocks that are frequently reused with revisions.

Developing a procedural layout generator involves codifying layout design procedures through parameterized program scripts. Similar to full-custom manual design flow, designers carefully consider the placement and routing of layout elements. However, unlike the manual design process that produces a single fixed layout, the generator design process creates a parameterized script capable of generating various layout instances. Each element's geometry, including boundaries, paths, and structure references, is defined by variables and formulas that change according to input parameters and design rules (Fig. 4 [4]). By describing geometric constraints as formulas, these generators can produce multiple layout variants without manual redrawing. This design approach allows designers to encapsulate their expertise into a program that quickly produces layout variants.

\subsection{Layout-to-Generator Conversion Framework}

\begin{figure}[t]
        \centering
        \includegraphics[width=\linewidth]{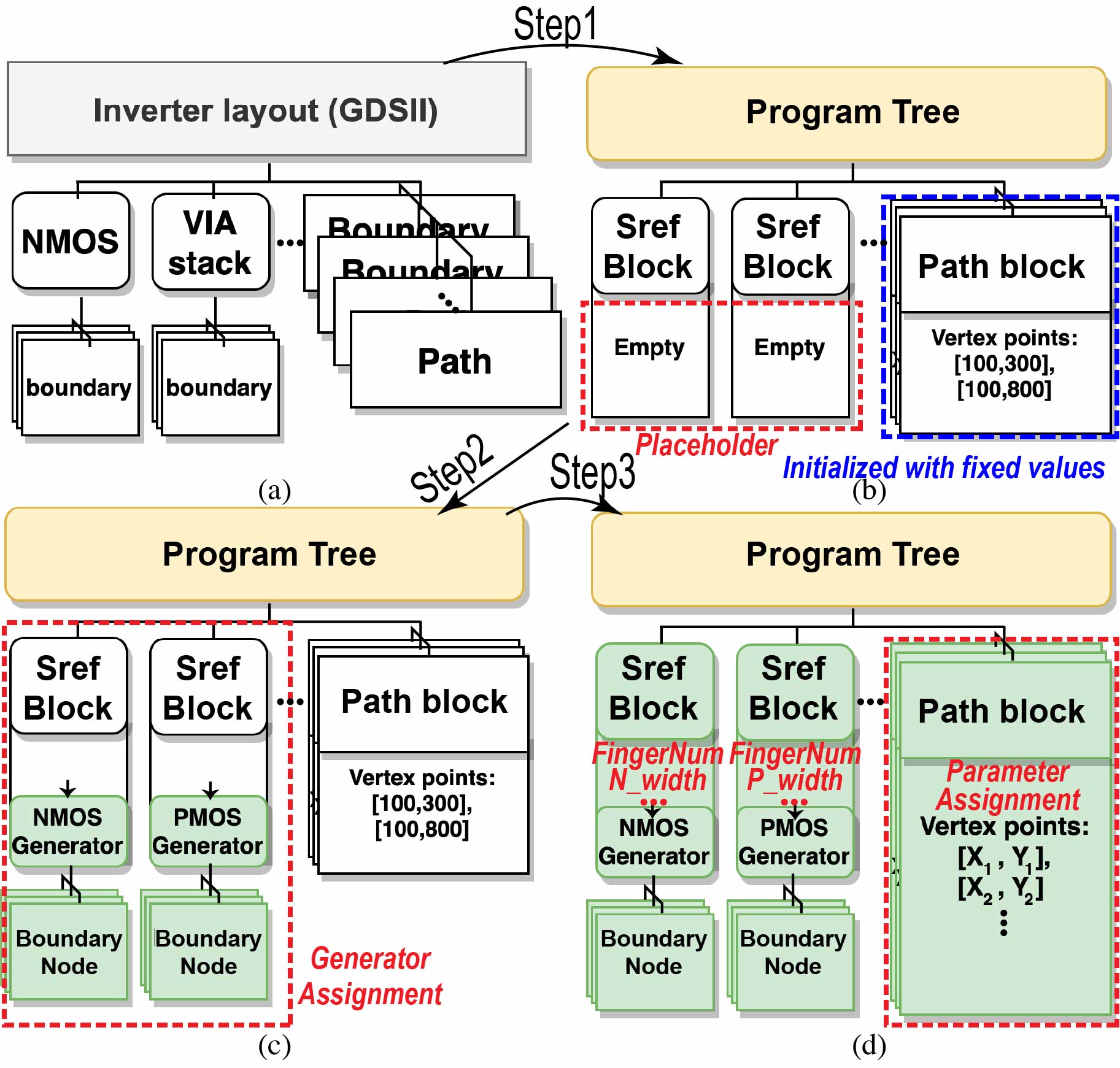}
        \caption{An example hierarchical design flow of the layout-to-generator conversion framework with inverter example, reprinted from [4].}
\end{figure}

The layout-to-generator conversion framework [4] is proposed to reduce the significant human effort required in developing procedural layout generators. Traditionally, developing a layout generator from scratch involves numerous manual steps, including defining layout elements, specifying design variables, describing parametric expressions, and programming the entire design procedure. This process is not only time-consuming but also requires human experts in both analog design and programming.

The layout-to-generator conversion framework addresses the challenges of traditional layout generator development by converting reference layouts into a generator. By referencing silicon-proved layout, this framework allows designers to use silicon-proved layouts as references. In addition, the framework reduces the manual coding workload by automating repetitive task. With this approach, designers with limited expertise in analog layout design or programming can develop high-quality layout generators.

The layout-to-generator conversion framework supports a hierarchical design flow to simplify the development process just as a layout cell can be hierarchically structured with its sub-cells. A layout cell, also simply called a cell, is a modular layout block containing elements like boundaries, paths, and smaller cells. The smaller cells instantiated within the larger cell are called sub-cells. If there are available generators that can generate the sub-cells in development of the generator of the larger cell, a designer can utilize them to create codes for generation of the sub-cells instead of building the complex layout generator of the larger cell from scratch. In the layout-to-generator conversion process, sub-cells can be converted to the generator codes using the sub-cell generators as illustrated in Fig. 5 [4]. This approach enables complex designs to be efficiently developed from simpler generators. By reusing existing sub-cell generators, designers can codify hierarchical structures without manually developing every generator from scratch. This approach reduces redundant coding effort and increases the efficiency of the layout-to-generator conversion process. 

However, despite the automation techniques provided by the layout-to-generator conversion framework in [4], several steps still require manual intervention. For example, in hierarchical design flow, for every sub-cell, a designer must manually check whether there is a generator that can generate the sub-cell and must assign the available generator to it. Additionally, describing parametric expressions still needs to be done manually. While the framework significantly reduces the overall workload through several automatic processes, further improvements can enhance the automation of the generator development process.

\section{Application Overview}

This work focuses on automating the assignment of generators to sub-cells, a critical part in layout-to-generator conversion for hierarchical designs. The proposed technique accelerates layout and generator development cycles, enhancing the efficiency of the layout-to-generator development workflow.

\subsection{Construction of A Large Layout Library}

\begin{figure}[b]
\centering
\includegraphics[width=\linewidth]{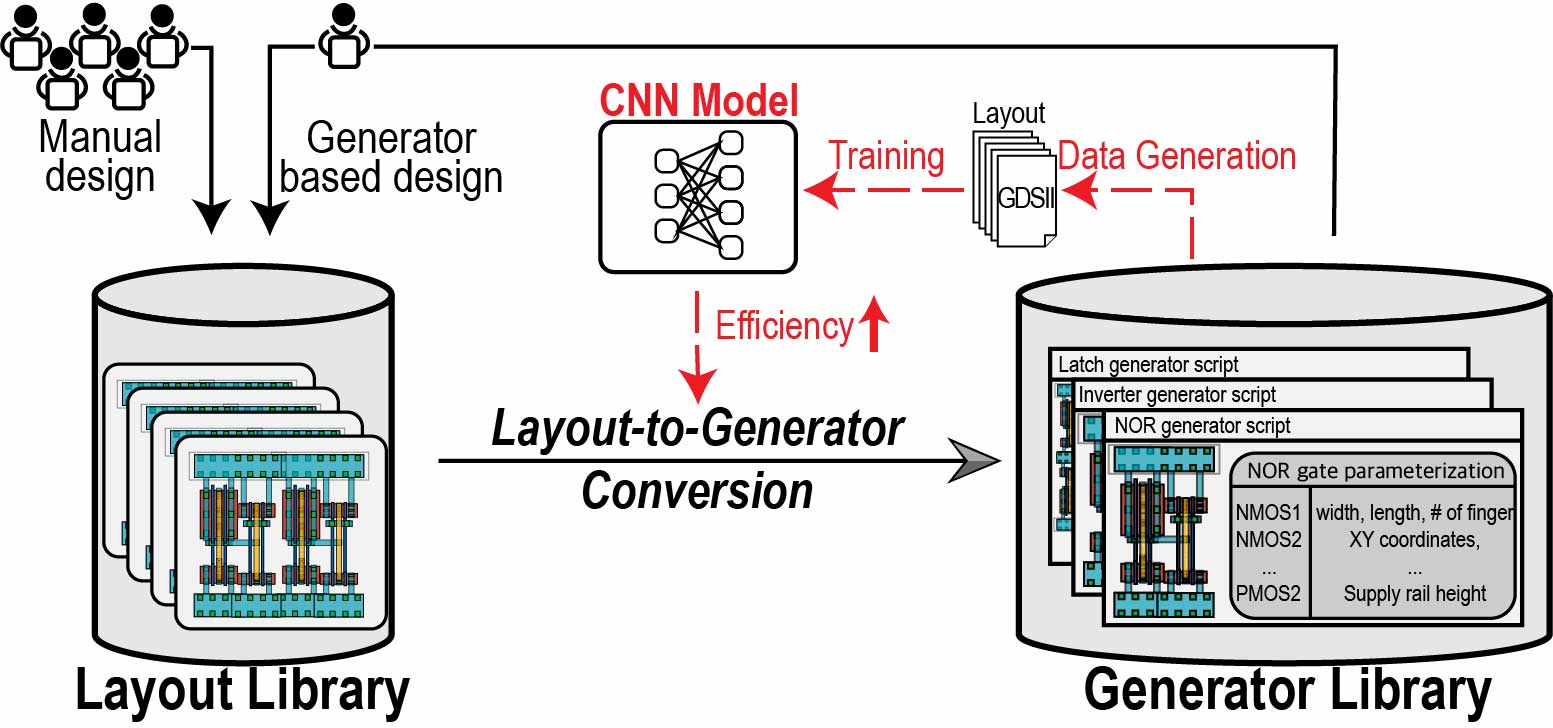}
\caption{A construction cycle of layout and generator libraries using a layout-to-generator conversion technique.}
\label{fig4_cycle}
\end{figure}

The proposed technique can be used to construct a large layout library for companies. Fig. 6 shows the concept of the construction cycle. While not fully automated, this cycle is supported by semi-automatic techniques to reduce manual workload, as described in [4]. This is an accumulative cycle that expands assets of layouts and layout generators by positive feedback. Both layout and generator libraries grow through this cycle while helping each other to grow. By layout-to-generator conversion, a designer can easily create layout generators from layouts in the layout library. By using the converted generators, the designer can quickly produce many new layouts. The libraries of layouts and generators grow with new assets. The designer can also develop new layouts by using the generated layouts as sub-cells. The new designs can be accumulated in the layout library. In addition, designers can manually create new layouts and add into the library during the cycle, increasing the diversity of circuit types and layout patterns. The newly developed layouts can be converted into new generators, and then the new generators can produce new layouts, again. By repeating this process, the layout and generator libraries continue to grow. 

The proposed CNN model also accelerates and evolves with the construction cycle. The CNN model assists in converting layouts to generators, and thus it helps the generator library become bigger. Because the CNN model is trained by layout data created by generators in the library, the bigger generator library improves the model. Because the model is improved, it further accelerates the layout-to-generator conversion and thus the growth of the generator library. Through this positive feedback cycle, the proposed CNN model evolves with the generator library.

\subsection{Layout-to-Generator Conversion Workflow}

\begin{figure}[b]
        \centering
        \includegraphics[width=0.9\linewidth]{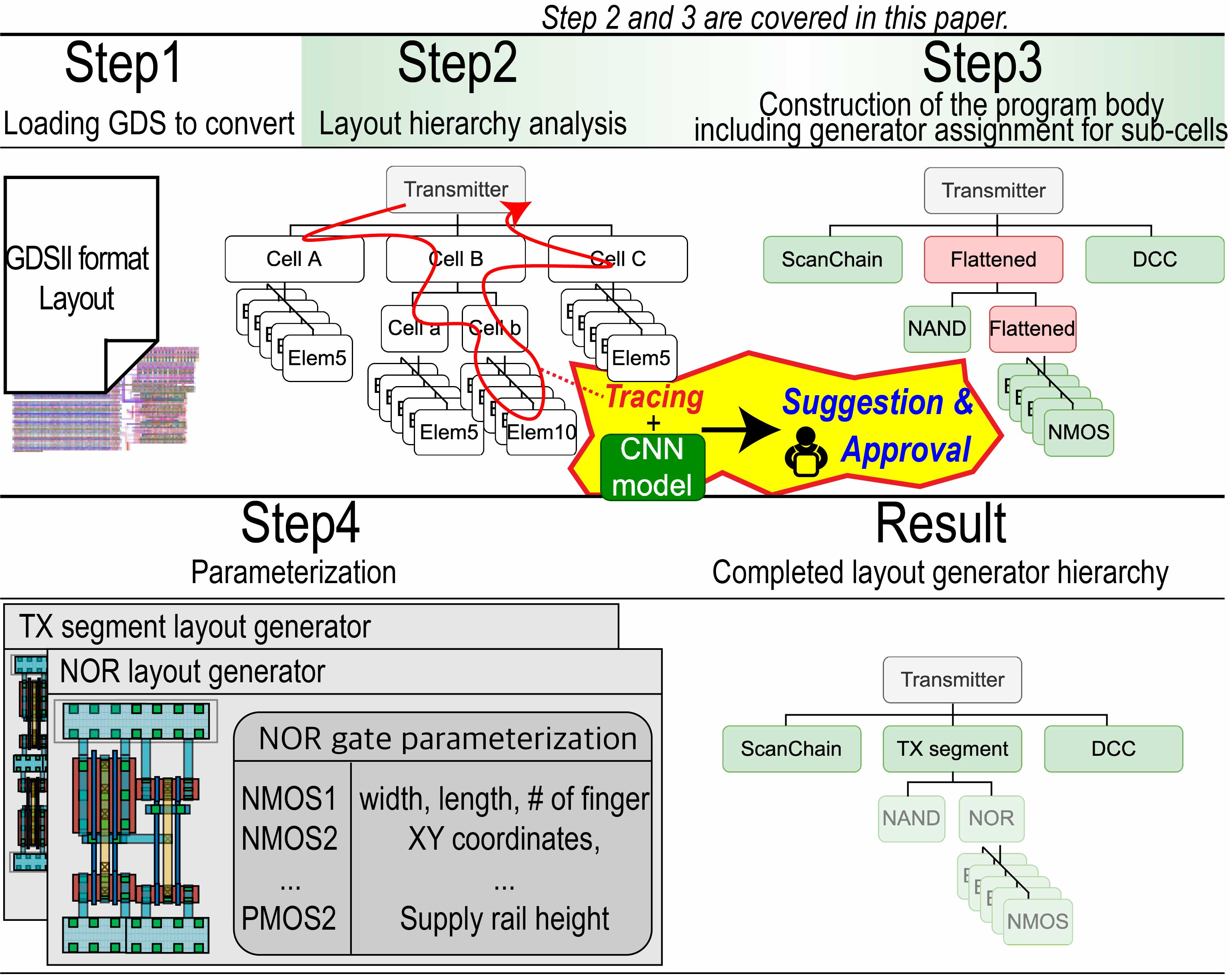}
        \caption{The workflow of layout-to-generation conversion assisted by the CNN model.}
        \label{fig5_flow}
\end{figure}
        
\begin{figure}[ht!]
        \centering
        \includegraphics[width=0.9\linewidth]{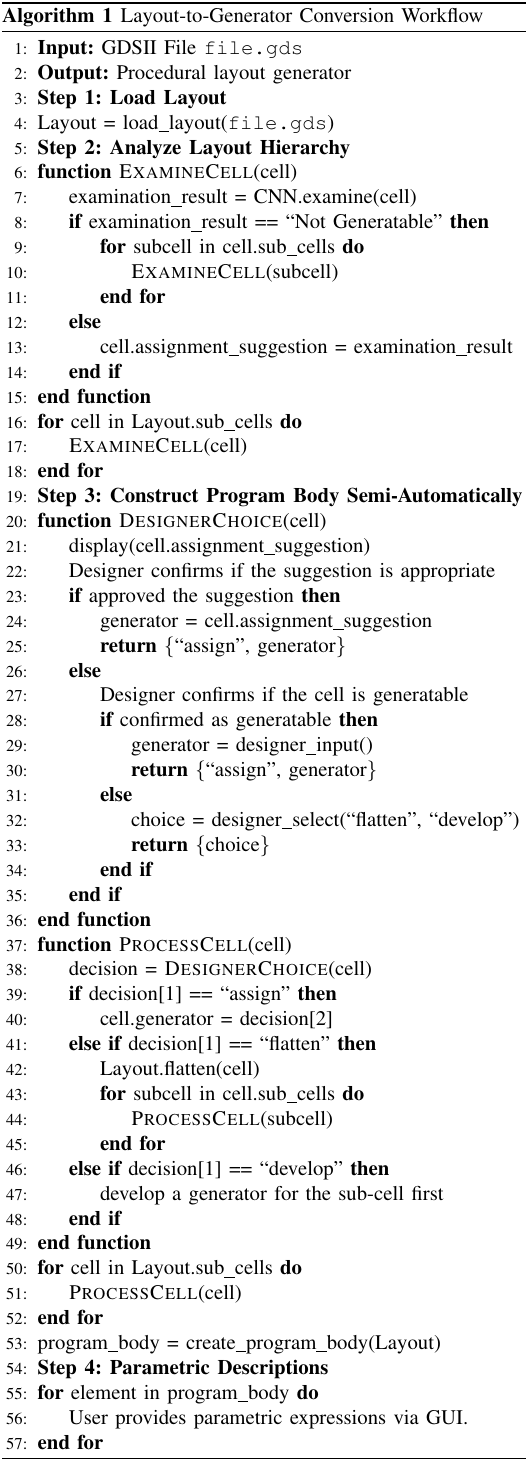}
        % \caption{The workflow of layout-to-generation conversion assisted by the CNN model.}
\end{figure}

Our proposed CNN model assists in the hierarchical design flow during layout-to-generator conversion process by automating the generator assignment for sub-cells, which was previously a time-consuming and error-prone step. Fig. 7 shows the workflow of the layout-to-generator conversion, while Algorithm 1 provides a detailed pseudo-code for each step. 

In Step 1, a reference layout is loaded. This initial loading prepares the data structure for subsequent hierarchical analysis.

In Step 2, the design hierarchy of the loaded layout is automatically analyzed in order to prepare for semi-automatic construction of the program body in Step 3. The program structure of the generator to be developed is planned in this step including generator assignment for sub-cells. While tracing the design hierarchy of the reference layout, sub-cells are encountered according to the hierarchical structure stored in the GDSII file. The CNN model then examines whether an encountered sub-cell can be generated by one of the generator scripts in the library. If the sub-cell is generatable, then the CNN model will automatically suggest using the correct generator script for the construction of the program body in Step 3. If not, the sub-cell can be flattened one hierarchy level and then the software traces down into the lower design hierarchy in order to continue to examine the flattened sub-cell. This iterative flattening process continues until a suitable generator is found. Or, the designer can choose to start developing the generator of the sub-cell, and then use it in Step 3. Our algorithm checks and memorizes the examination history in order to avoid unnecessary multiple examinations of the same sub-cell layout instance that is used multiple times in the reference layout design.

In Step 3, the software semi-automatically constructs the body of the generator program based on the analysis in Step 2. The skeleton of the program body is automatically prepared, and then generator code for sub-cells and basic elements (boundaries and paths) are inserted into the appropriate positions as planned in Step 2. In this process, the CNN model suggests the right generator scripts for the sub-cells if the sub-cells are generatable. Upon the designer’s approval, the suggested generator script is inserted in the right position. By reusing the generator scripts in the program body, a large amount of workload can be reduced. If there is no available generator for the sub-cell, the sub-cell can be flattened by one hierarchical level, or the designer can choose to develop the generator of sub-cell as explained previously. This suggestion-approval approach prevents a small chance of the wrong assignment of generator scripts. Because the CNN model cannot guarantee 100\% accuracy, it may mistakenly suggest the wrong generator scripts. Because the wrong usage of the generators can cause very complicated design rule violations or LVS violations, it must be prevented. Even with the possibility of occasional misclassifications by the CNN model, the overall workload of manually examining each sub-cell is greatly reduced.

In Step 4, the designer fills key parametric descriptions in the program body prepared in Step 3. This step requires manual GUI operation: input, but the GUI-based framework [4] helps designers describe parametric expressions through an GUI. Finally, the designer has the layout generator converted from the reference layout. The converted generator can produce various layouts in the same style of the reference layout but with different design parameters.

\section{Proposed Method}
In this section, we present our proposed method in order to automate sub-cell generator assignment using a CNN. We will explain our proposed method in order of data preparation for the CNN model, model architecture design, and model training.

\subsection{Data Preparation and Analysis}

\subsubsection{Data Preprocessing}

In preprocessing, to feed sub-cell layout data into the CNN model, the sub-cell layouts in a GDSII format are converted into matrix inputs whose elements are numeric values. The converted matrix inputs have 21 channels, each of which looks like a bitmap image matrix (Fig. 8) \cite{well}.

\begin{figure}[t]
        \centering
        \includegraphics[width=0.9\linewidth]{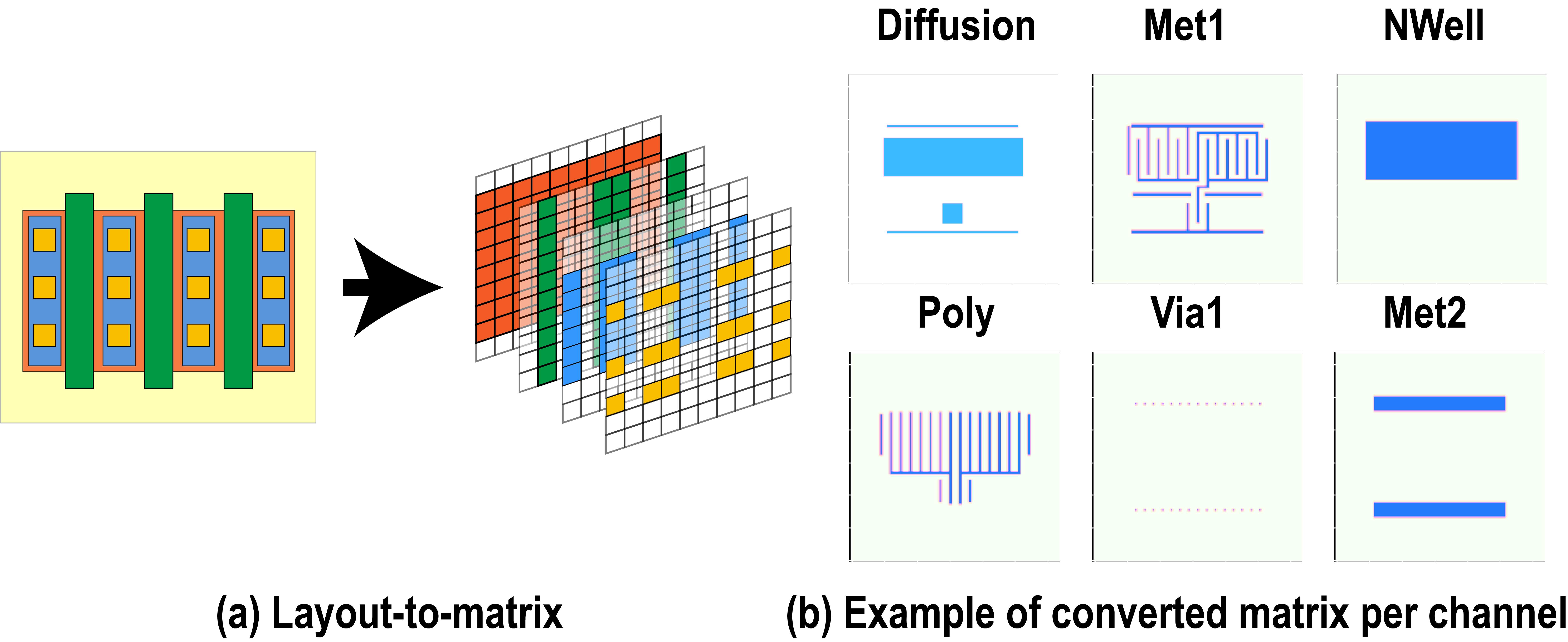}
        \caption{Preprocessing of a layout input.}
        \label{fig6_cnn}
        \end{figure}

This conversion involves three primary steps: 1) dividing a layout by physical layers, 2) converting each physical layer to a matrix input for each channel, and 3) re-sizing the matrix input sizes to 256x256.

In the first step, the layout input is divided by physical layers in order to preserve physical layer information. Consequently, the proposed CNN model differs from typical image classification models in the number of input channels. A typical color image has three different channels for R, G, and B whereas our CNN model uses 21 channels.  Each channel represents a specific physical layout layer, such as ‘metal1’, ‘metal2’, ‘poly’, ‘contact’, ‘via’, and ‘doping layer’. This approach enables the model to distinguish elements of different physical layers. To handle various layout designs in different technology nodes, similar functional purpose layers in different technology nodes are appropriately mapped to the same channel. For example, to draw gates of transistors, a poly-gate layer is used in old technology while a metal-gate layer is used in recent technology. Although the material and technology modes are different, both layers are used for the transistor gate region. Therefore, the poly-gate and metal-gate layers are mapped to the same channel.  

In the next step, we convert the format of the layer input of each channel to a matrix form from the GDSII format because the matrix format is suitable for a CNN model. For conversion, a grid is drawn on a physical layer design of the channel to define pixels for the design. For each channel, if a pixel overlaps with the elements drawn on the physical layer then the pixel value is set to 1. Otherwise, the pixel value is set to 0. For missing physical layers that were not used in the input layout, the corresponding channels are zero-padded. Through this process, the physical layer input to each channel is converted into a matrix format. 

In the last step, the matrix of each channel is resized to the same size in order to ensure that the input to the CNN model has the same dimension. For example, if the desired input matrix size is 256x256 then the resizing process works as follows. If the size of the converted matrix is 256x256 then no resizing process is required. If the converted matrix is smaller or larger than 256x256, then zero-padding or average sub-sampling is used to resize the input matrix to 256x256, respectively. 

\subsubsection{Cell Size Distribution}

To determine the appropriate input matrix size, we analyzed the size distribution of the sub-cells. We studied how improper matrix size can lead to poor performance of the model. If the input matrix size is too small (low-resolution), many sub-cells lose geometric information by excessive downsampling during converting layout data into matrices, and thus reducing the model’s accuracy. Conversely, an excessively large input matrix size would make a model less efficient due to too much computation. 

\begin{table}[t]
        \caption{Sub-cell Size Distribution}
        \label{tab1_distribution}
        \centering
        \includegraphics[width=\linewidth]{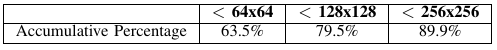}
\end{table}

We determined the appropriate input matrix size by obtaining the size distribution of the sub-cells. For computational efficiency, candidates of input size were restricted to powers of two (32x32, 64x64, 128x128, etc.). Among 11,539 sub-cells, only 36.5\%, 20.5\%, and 10.1\% sub-cells are larger than 64x64, 128x128, and 256x256, respectively (Table I). Therefore, if the input matrix size is 256x256 (equivalently 2,560 nm x 2,560 nm area), then 89.9\% of the sub-cells out of the 11,539 sub-cells can be handled by the input matrix without resizing. Only about 10.1\% of sub-cells lose information during data preparation if the input matrix size is 256x256. In addition, increasing the input matrix size beyond 256x256 would lead to a significant increase in memory requirements. Therefore, because of the diminishing return beyond the matrix size of the 256x256, we designed the input matrix size of our CNN model to 256x256.

\subsection{Model Architecture}
We adopted a convolutional neural network (CNN) to examine whether a sub-cell can be generated by generator scripts available in the library. The CNN model classifies a sub-cell layout into one of 52 classes depending on how the available generators in the library can be utilized in generating the input sub-cell. 51 classes represent 51 scenarios where the sub-cell layout input can be generated by one of the available generator scripts whereas one class is labeled as ‘Not generatable’ to indicate that the input sub-cell cannot be generated by any available generator script. The choice of 52 classes is based on the available generators at the time of development. This number is not fixed to a particular technology node and can be easily adjusted if more generator scripts are added or removed in the future. Our tool infers that the input layout cannot be generated by an available script if the not-generatable class has the maximum probability or no other probability is larger than a threshold level. Otherwise, it infers that the generator associated with the maximum probability output can generate the input layout.

\begin{figure}[t]
\centering
\includegraphics[width=\linewidth]{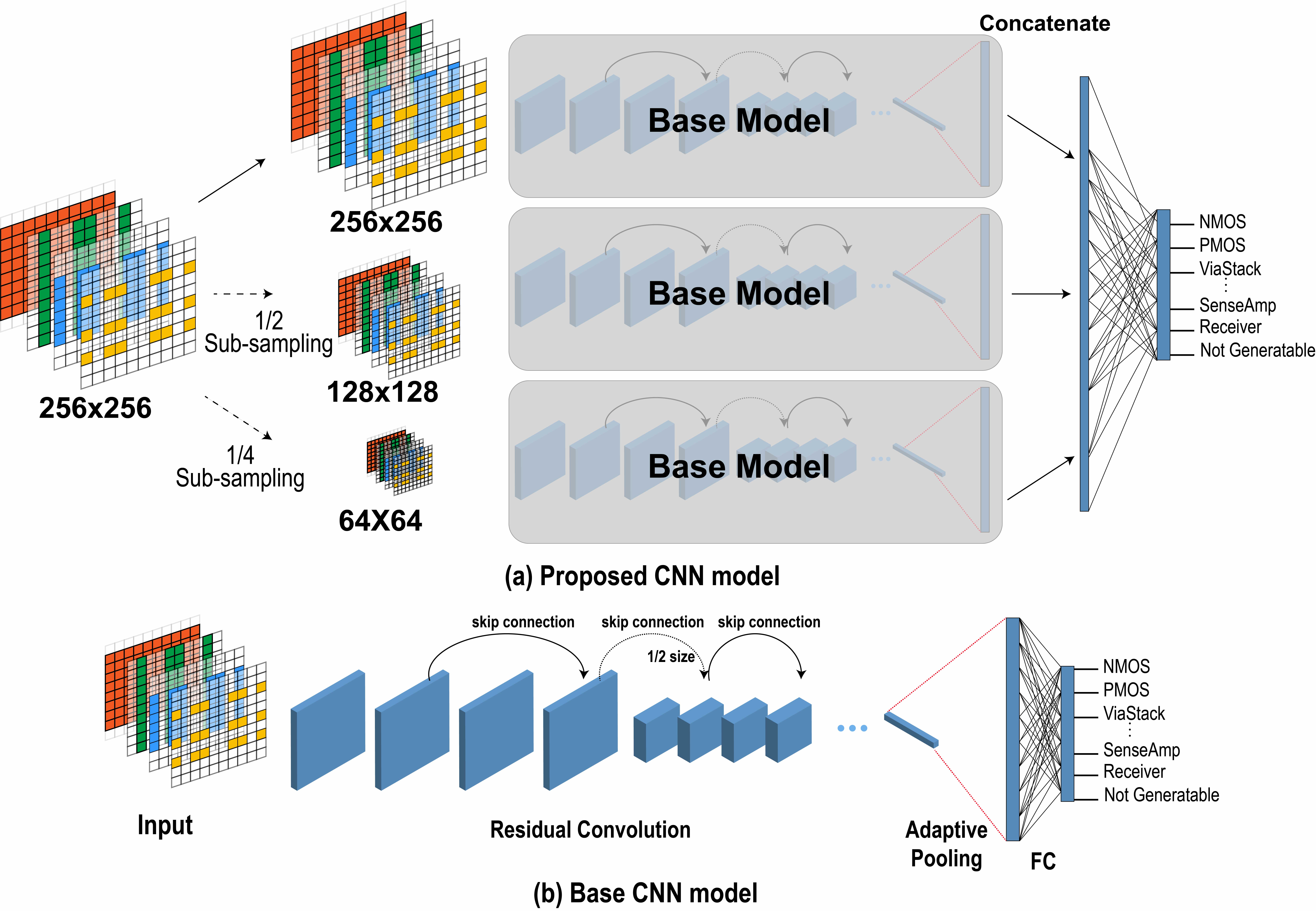}
\caption{The architectures of the CNN models (a) the proposed CNN model, (b) the base model.}
\label{fig7_cnn}
\end{figure}

The proposed CNN model is described in Fig. 9a. The proposed CNN model (Fig. 9a) is modified from the base CNN model (Fig. 9b) which is very similar to ResNet \cite{res}. The base CNN model is different from a typical ResNet in fact that the number of input channels is 21 instead of 3. The proposed CNN model combines three base CNN models with three different input matrix dimensions of 64x64, 128x128, and 256x256 for multi-scale processing (Fig. 9a). 

The base CNN model consists of residual blocks with convolutional layers and skip connections, pooling layers, and fully-connected layers like a typical image classification CNN model (Fig. 9b) \cite{res}. The base CNN model receives an NxNx21 input. Each residual block contains two convolutional layers with 3x3 kernels, with batch normalization and ReLU activation layers. Skip connections for each residual block connect the input of the residual block directly to its output. The last fully-connected layer with 52 nodes was used at the end to classify the input sub-cell layout into one of 52 classes of generation scenarios. The output of the last layer is a vector of 52 elements corresponding to the probability of 52 classes. Among the 52 classes, one class is the not-generatable class which indicates that no available generator can generate the input sub-cell.  
 
We used a sigmoid function at the last layer instead of a softmax function, which is widely used in multiclass classification problems \cite{sig1}, \cite{sig2}. In the applications of softmax, the sum of the output probabilities is always ‘1’ assuming that an input always belongs to one of the classes. Consequently, softmax tends to assign high probability to the most likely class, even when the input layout does not correspond to any generator. Therefore, the ‘Not generatable’ class must be considered one of the outputs if a softmax is used. However, layouts that cannot be generated by the available generator scripts are too diverse to be treated as one class ‘Not generatable’ in our experience. In contrast, sigmoid functions evaluate each class independently, assigning probabilities between 0 and 1. Among these probabilities, The model selects the class with the highest probability if it exceeds a predefined threshold. If no class exceeds the threshold, or if the 'Not generatable' class scores the highest, the input sub-cell is classified as not generatable. This approach helps the model handle diverse not-generatable layouts. Therefore, we adopted sigmoid functions and treated the task as multiple binary classification problems, where the probability of each class is independent from others

The proposed CNN model utilizes three base models in parallel sub-network in order to support a multi-scale approach by extracting three different scale feature maps (Fig. 9a). The proposed model begins with an input matrix of size 256x256. This matrix undergoes sub-sampling at two different ratios of  4:1 and 2:1, producing matrices of sizes 64x64 and 128x128, respectively. The two produced matrices (64x64 and 128x128) and the  input matrix (256x256) are fed into the three sub-networks, all of which have the same depth and architecture. Each sub-network architecture is identical to the base model without the last fully-connected layer. After three matrices are processed by the three sub-networks, the outputs from the sub-networks are concatenated and then fed to the last fully-connected layer to classify the generation probabilities.

Sub-networks with different input matrix sizes in the proposed CNN model improve accuracy by extracting and utilizing multi-scale feature maps. Because of down-sampling, the sub-network with the small matrix input (64x64) is advantageous in recognizing large structures of the input layout such as placement of its sub-cells and long routing patterns between them. However, this sub-network may lose details of the layout such as local routing patterns, via connection, etc., due to excessive down-sampling. On the other hand, the sub-network with the large matrix input can preserve more detailed features, such as intricate routing patterns and fingers of gates. Similarly, the sub-network with intermediate input matrix size (128x128) is advantageous in recognizing structures and patterns of intermediate sizes. By incorporating multiple sub-networks with different input sizes, the accuracy of the proposed CNN model is improved by up to 4.5\% in the test.

\begin{table}[t]
        \caption{DEFINITION OF A CONFUSION MATRIX FOR GENERATABLE V.S. NOT-GENERATABLE INPUT LAYOUTS}
        \label{tab2_confusion_matrix}
        \centering
        \includegraphics[width=\linewidth]{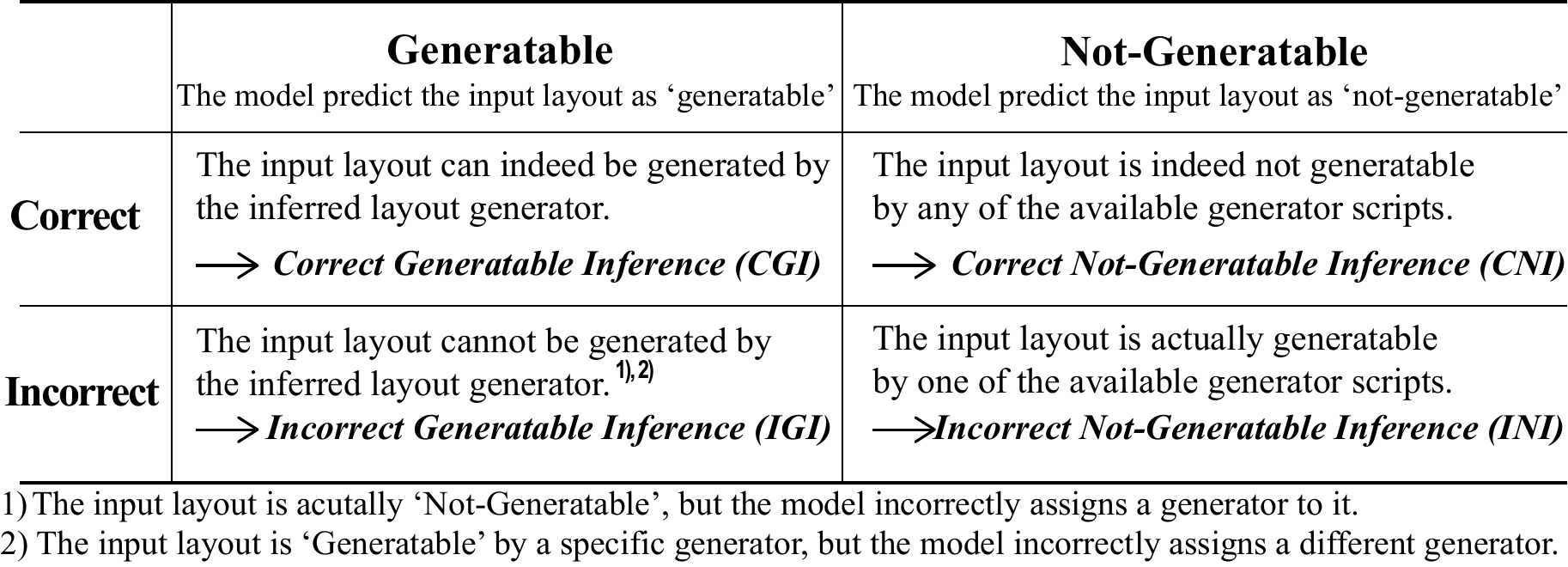}
\end{table}

Hyperparameters such as depth of the neural network, matrix dimensions, and the number of layers were manually tuned by observing metrics while training the model. The relevant metrics are defined based on the confusion matrix presented in Table II.

Precision is defined as the ratio of correctly identified generatable instances (CGI) to the total number of generatable inferences, including incorrect ones (IGI). 
\begin{equation}
        \text{Precision} = \frac{\text{CGI}}{\text{CGI} + \text{IGI}}
\end{equation}
Recall is defined as the ratio of correctly identified generatable instances (CGI) to the total CGI and missed generatable instances (INI).
\begin{equation}
        \text{Recall} = \frac{\text{CGI}}{\text{CGI} + \text{INI}}
\end{equation}

Accuracy provides an overall measure of the model’s classification performance. 
\begin{equation}
        \text{Accuracy} = \frac{\text{CGI} + \text{CNI}}{\text{CGI} + \text{CNI} + \text{IGI} + \text{INI}}
\end{equation}

The F-0.5 score is a weighted harmonic mean of precision and recall, with more emphasis on precision.
\begin{equation}
        \text{F-0.5 Score} = \frac{(1 + 0.5^2) \times (\text{Precision} \times \text{Recall})}{0.5^2 \times \text{Precision} + \text{Recall}}
\end{equation}

Not-Generatable Identification Rate (NGIR) focuses on the model’s capability to identify layouts that cannot be generated. It is calculated as the proportion of correct not-generatable inferences (CNI) to the total of CNI and incorrect not-generatable inferences (INI).
\begin{equation}
        \text{NGIR} = \frac{\text{CNI}}{\text{CNI} + \text{INI}}
\end{equation}

\subsection{Training}

In order to train the model, we used three different types of datasets. 1) a dataset of layouts was generated with random parameters using available generators in the library. 2) As negative samples, not-generatable layouts were also randomly created and used. 3) We also used manually-designed silicon-proved layouts as a realistic dataset \cite{moon}. All training datasets were generated or designed in a 28 nm CMOS technology node.

\begin{figure}[t]
\centering
\includegraphics[width=\linewidth]{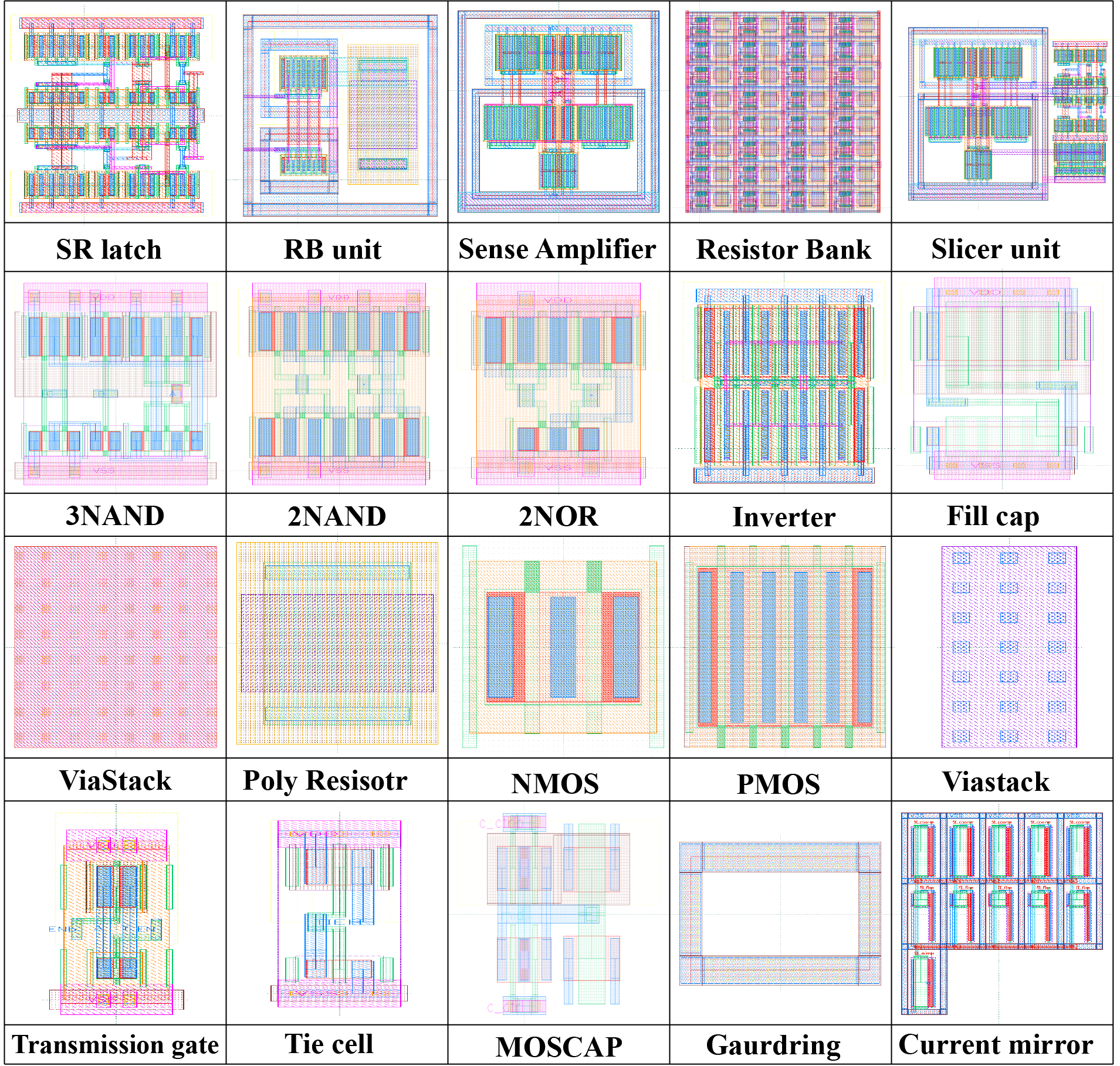}
\caption{Example generatable sub-cell layouts that were correctly classified  by the proposed CNN model. Note that the CNN model must be able to distinguish subtle differences in layout styles in order to determine whether a layout can be generated by a generator template.}
\label{fig9_layout_exs}
\end{figure}

The first dataset was acquired by randomly generating layouts with generator scripts available in the library. The generator scripts include generators for boundary arrays, via stack arrays, MOSFETs, poly resistors, 2-input NAND gates, 3-input NAND gates, transmission gates, current mirrors, inverters, high-speed set-reset (SR) latches, sense amplifiers, resistor banks, and so on (Fig. 10). For generation of dataset, input parameters of the generator scripts were randomly generated within the acceptable ranges defined by the developers. Using the randomly generated parameter sets, various layout datasets were generated. For this automated process, the designer specified the layout generators, defined parameter ranges, and set the number of samples. The algorithm then randomly selected parameters, executed the generators, and converted the resulting layouts into matrices for preprocessing.

To address a data imbalance problem, we added a dataset of randomly created not-generatable layouts. As negative sample layouts, several boundaries were randomly created: they were randomly sized and randomly located on random layers. These negative samples are meaningless random layouts that cannot be generated by the available generators, but improves the model’s accuracy. Similarly, this process of creating not-generatable layouts was also automated, requiring only specifying the number of samples.

By training the model with these negative samples, we could mitigate the problem of bias from imbalanced data. The layouts that are generated by available generators do not include not-generatable layouts, and thus training the model only with them causes a data imbalance problem: the imbalanced data with only generatable samples can bias the model toward inferring generatable classes, leading to fewer inferences of the not-generatable case. 

To add realistic cases to training data, we also used manually-designed layouts acquired from a fabricated and tested chip \cite{moon}. We manually labeled layouts of a high-speed wireline transmitter design, and used them for training. The transmitter design was developed independently of this work, and thus these data are not biased by our intention. This dataset includes both realistic generatable and not-generatable layouts.

The dataset of manually-designed layouts improves the generalization ability of the model by increasing the variety of realistic design samples. With improved generalization ability, the model performs well on unseen data that are not included in the training data. To improve the generalization ability, it is important to use realistic negative samples in training so that the model can distinguish the subtle difference between generatable and not-generatable layouts in real design. In practical design cases, there are many not-generatable layouts that look very similar to generatable ones. For example, an inverter and a transmission gate may have almost identical placement of transistors, but their routing patterns are slightly different. The subtle differences between these can hardly be trained by the randomly created negative samples which do not look like not-generatable layouts in real design. Although the randomly created not-generatable layouts in the second dataset help address the data imbalance problem, they do not provide such realistic negative samples. Because the third dataset of manual layouts is acquired from a real chip design, they provide realistic samples for both positive and negative cases. These samples help the model to identify tricky differences in layout styles improving the generalization ability of the model.

The model was trained with 25,500 layouts that were generated by available generators, 500 not-generatable layouts that were randomly created, and 442 manually-designed layouts.

In order to avoid overfitting and also for good generalization, the model was validated using manual layouts acquired from another chip design different from the one used for the training dataset. During training, for validation, we used 1,888 manual layouts that were neither used in training nor testing. The validation dataset was designed also in a 28 nm CMOS technology node. For the training, we used the binary cross entropy loss function and selected a batch size of 8. Training continued until the validation loss showed no improvement for a predefined number of epochs, thereby preventing overfitting.

We trained the model to classify sub-cell layouts into one of 52 classes (Fig. 10). The 52 classes comprise 22 different various circuit layout classes, 28 different via stack array classes, 1 boundary array class, and 1 not-generatable class. It is noticeable that 28 different via stack array classes can be generated by one via stack layout generator with various different layer combinations. We found that a human designer usually has to spend a lot of time checking the layer options of via stacks to utilize generators correctly because there are many various layer options of via stacks. In order to reduce the time required to manually check via stack layer options, we designed the CNN model to classify the vias stack options.

\section{Experimental Result}
We evaluated the performance and efficiency of the CNN model through three experiments: 1) assessing classification accuracy of the proposed CNN, 2) measuring runtime of the proposed CNN, and 3) analyzing and comparing the accuracy metrics and runtimes of the proposed and base CNN models.

\begin{figure}[t]
\centering
\includegraphics[width=\linewidth]{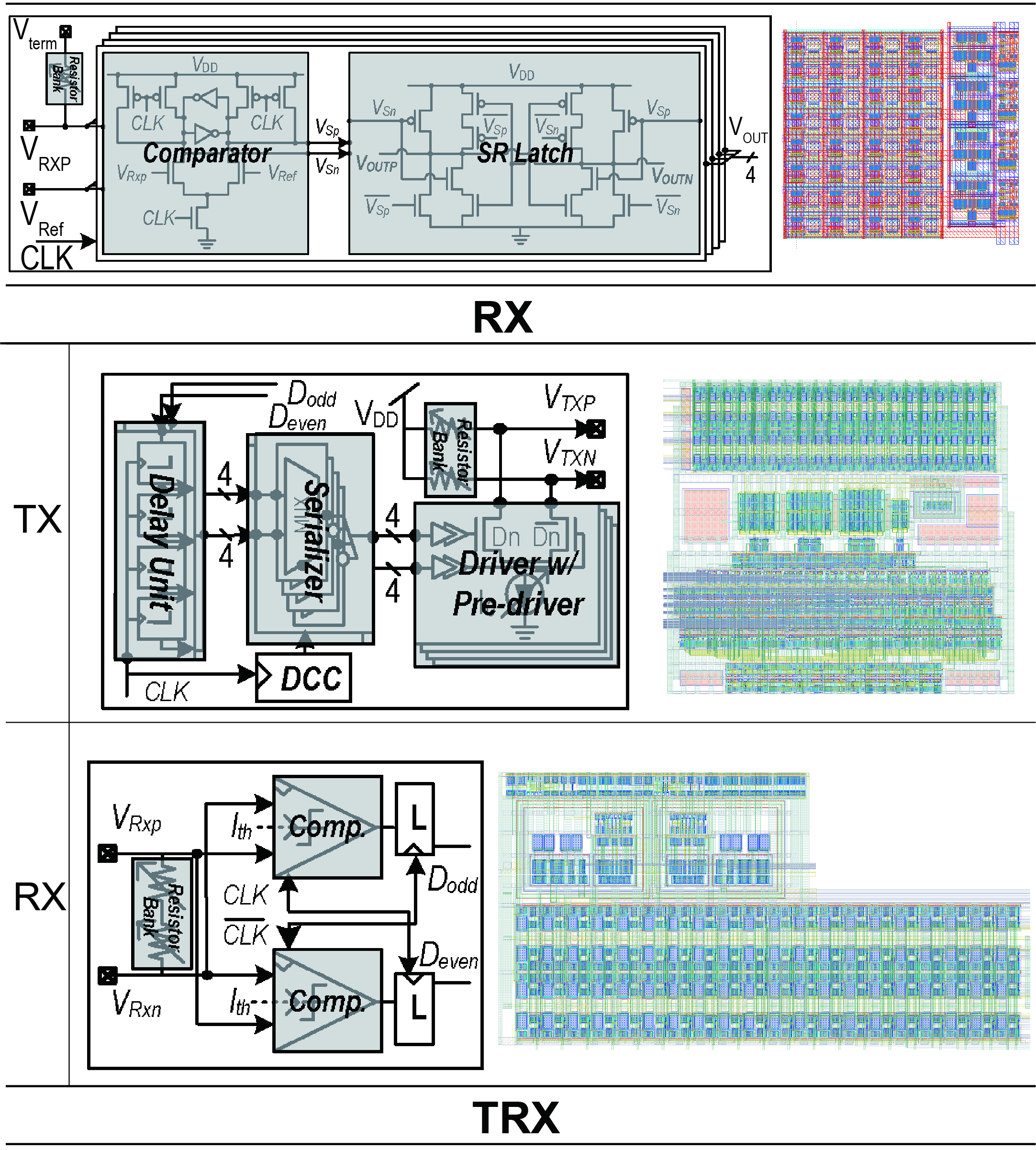}
\caption{Tested high-speed I/O circuits: (a) a compact quarter-rate RX and (b) a TRX block with relaxed impedance matching capability.}
\label{fig10_circuits}
\end{figure}

The model was tested with two manually-designed high-speed I/O circuits (RX \cite{rx}, TRX \cite{trx}) (Fig. 11). The RX circuit, implemented in a 28 nm CMOS process, is a compact single-ended receiver that operates at 20 Gbps for short-reach links. The TRX is implemented in a 65 nm CMOS process. Its transmitter uses a four-tap feed-forward equalization in half-rate architecture and configurable resistor banks to accommodate different impedances.

For each circuit, a dataset was prepared by converting its sub-cells into matrices. RX contains only generatable sub-cells. The generators of all RX’s sub-cells were used to prepare the training dataset, among others. Therefore, all sub-cells of RX must be inferred as “generatable”, making the dataset retrieved from RX the easiest one for the model. TRX was designed in 65 nm CMOS technology and contains not-generatable sub-cells. Recognizing the sub-cells of TRX is more challenging than the ones of RX for two reasons: 1) the technology node of TRX is different from the ones of the training dataset, 2) the layout patterns of not-generatable sub-cells of TRX are unfamiliar to the model because these layouts were not included in the training dataset. 

\subsection{Classification Accuracy}
\begin{table}[t]
        \caption{DETAILED INFERENCE RESULTS BY THE TYPE OF GENERATOR SCRIPTS.}
        \label{tab2}
        \centering
        \includegraphics[width=0.9\linewidth]{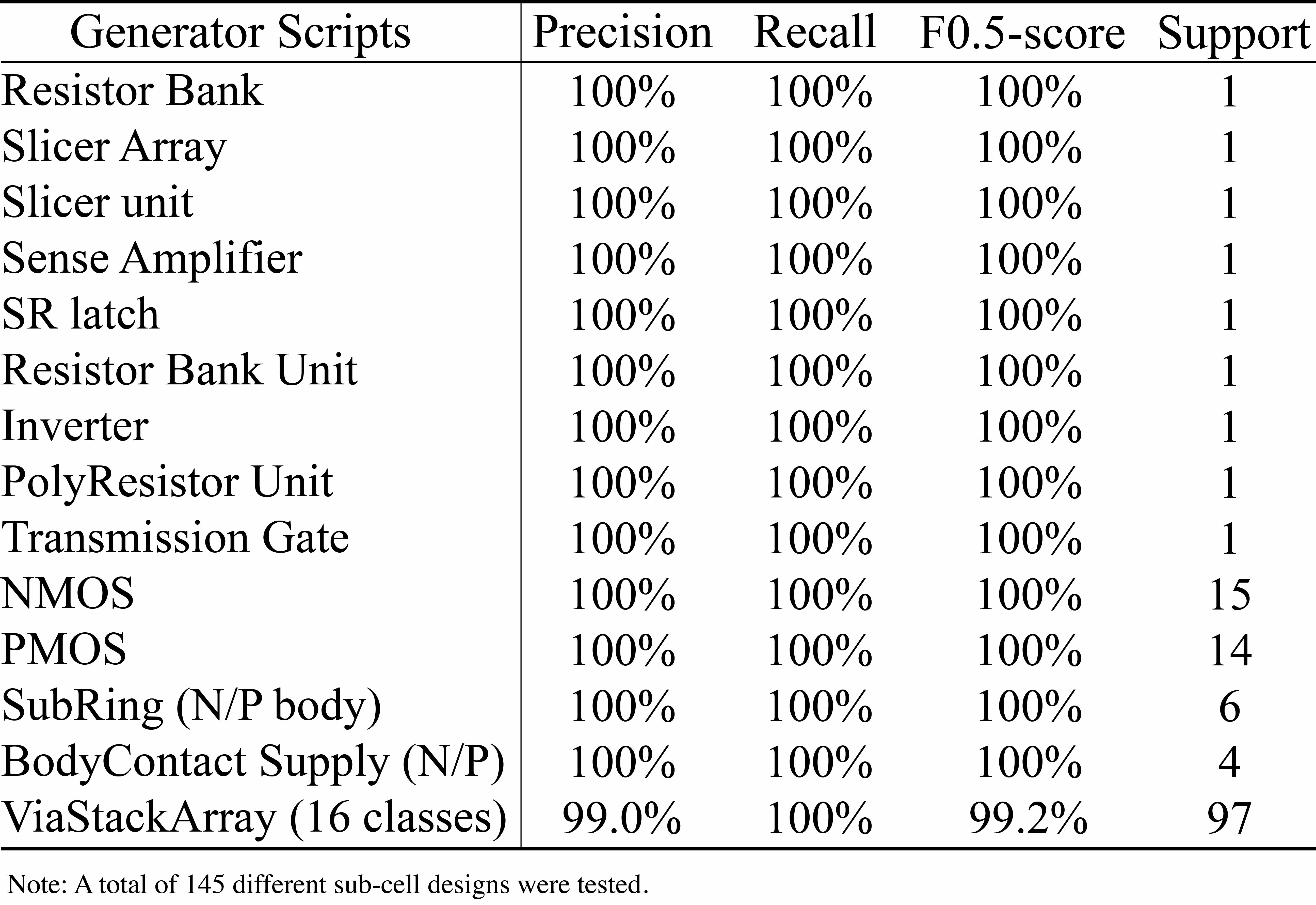}
\end{table}

\begin{figure}[t]
\centering
\includegraphics[width=0.9\linewidth]{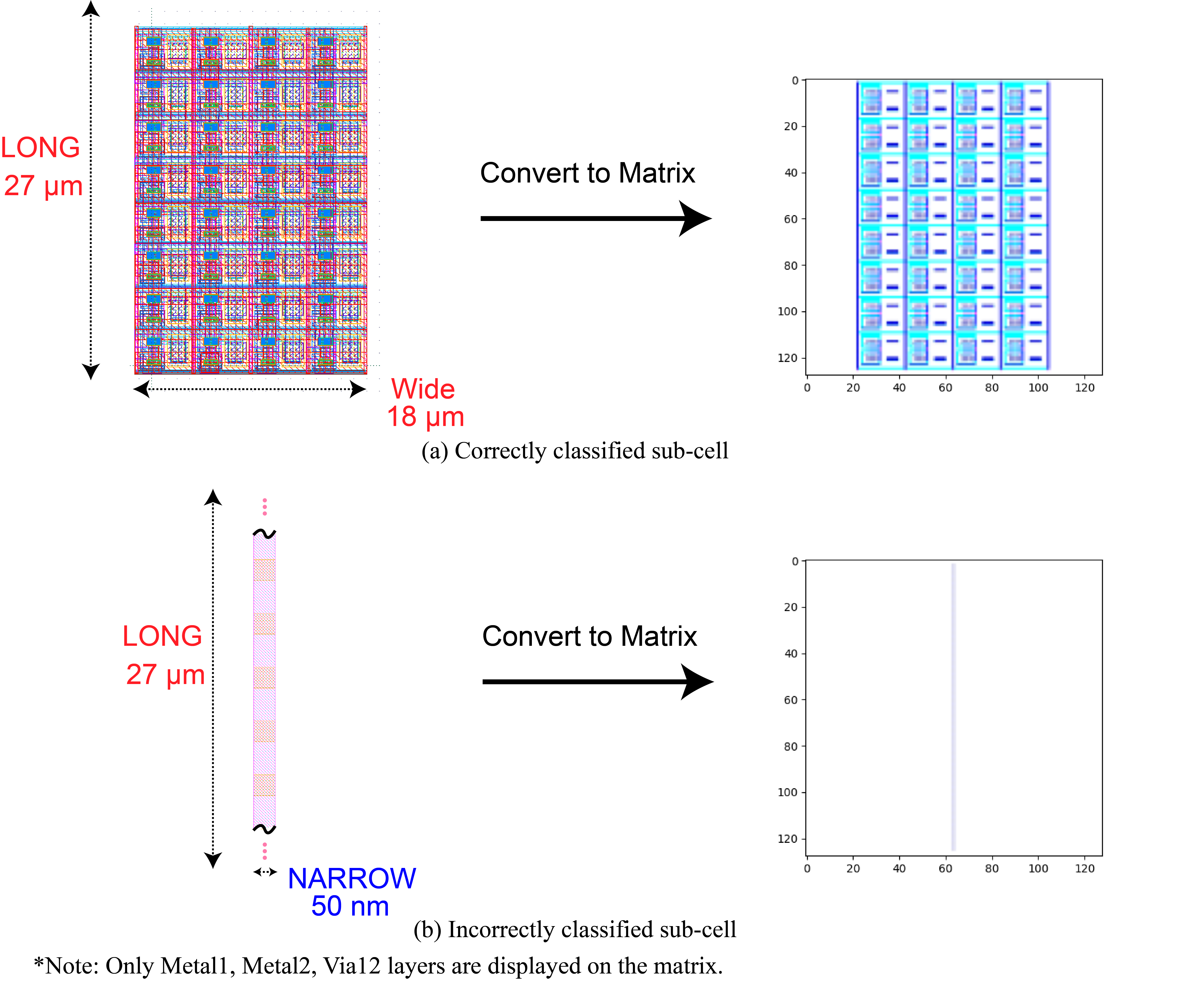}
\caption{Example layouts that were (a) correctly and (b) incorrectly classified; A layout with an extremely poor aspect ratio was incorrectly inferred by the CNN model.}
\label{fig11_fail}
\end{figure}

To evaluate the performance of the proposed CNN model under diverse scenarios, we tested the trained CNN model with the two different circuits RX and TRX. 

First, we tested the model with the RX circuit to evaluate the model’s performance on data similar to the training dataset. We recorded the examination count of sub-cells and the inference results. Identical sub-cell layouts are counted only once in the number of examinations because our algorithm avoids multiple examinations of the identical layout instances referencing the same design source by checking examination history. As a result, the same generator was assigned once for the identical layouts.  Our software examined only 145 different sub-cells instead of examining all 4,885 sub-cells.

Table III shows the detailed inference results by each type of generator script when the model is tested with the RX. Because a total of 4,885 RX’s sub-cells were instantiated by referencing 145 different sub-cell designs, the model examined only the 145 different sub-cell designs instead of examining 4,885 instances. 144 sub-cell designs were correctly classified into 31 different classes out of a total of 52 classes. Only one sub-cell design was incorrectly classified: this is a vertically long via stack array layout with extreme aspect ratio (Fig. 12). It seems that this sub-cell is too long and too narrow to be identified even at the highest resolution of 256x256. To classify this large sub-cell correctly, higher resolution would be necessary. Although one sub-cell design was incorrectly classified, the incorrect generator assignment for the sub-cell can be prevented by suggestion-and-user-approval process of our tool flow. Other sub-cells were classified correctly. In overall, the model achieved 99.3\% precision, 100\% recall, 99.4\% F-0.5 score, and 99.3\% accuracy. This result implies that the proposed method can successfully suggest generator scripts for the sub-cells for layout-to-generator conversion.

\begin{table}[t]
\caption{CORRECT AND INCORRECT TABLE RESULTS FOR TRX\newline
(A) TOP-1 PREDICTION, (B) TOP-3 PREDICTION.
}
\label{tab3}
\centering
\includegraphics[width=0.8\linewidth]{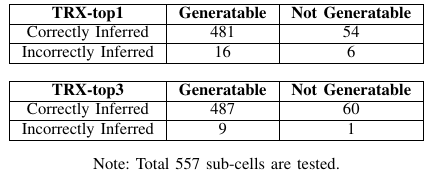}
\end{table}

Secondly, we evaluated the model’s performance using the TRX, a high-speed I/O transceiver that was manually designed in 65 nm CMOS technology. This dataset is particularly challenging because the model was trained with a different technology node (28 nm CMOS technology) and TRX contains 69 not-generatable sub-cells that is not familiar to the model. TRX has a total of 17,206 sub-cell instances, referencing 557 different sub-cell designs. Among 557 sub-cell designs, 69 sub-cell designs were not generatable. The model successfully classified 535 sub-cell designs out of 557, achieving accuracy of 96.1\% (Table IV.a). Moreover, the model achieved a top-3 accuracy of 98.2\%, meaning the correct prediction was among the model's top 3 most confident predictions in 547 of 557 sub-cells (Table IV.b). By providing the list of the top-3 most confident predictions during the designer approval stage, the chance of correctly suggesting the generator assignment to the designers can be greatly improved. In overall, the model achieved an F-0.5 score of 97.2\% and successfully identified unfamiliar layouts as not-generatable sub-cells, such as duty cycle corrections (DCC)s, pre-amplifiers, switches, delay units. These results demonstrate that the model can classify unfamiliar sub-cells designed in a different technology node.

In a practical scenario for layout-to-generator conversion, the CNN model’s accuracy would be higher than the second test result (TRX) for three reasons: 1) the model classifies familiar layouts well. 2) there is a high chance that small sub-cells are familiar to the model. 3) the majority of sub-cells are small sub-cells that can be easily classified by the CNN model.

\begin{table}[t]
        \caption{COMPARISON OF SUB-CELL EXAMINATION TIMES BY THE PROPOSED MODEL AND BY MANUAL CHECKING.}
        \label{tab4}
        \centering
        \includegraphics[width=0.85\linewidth]{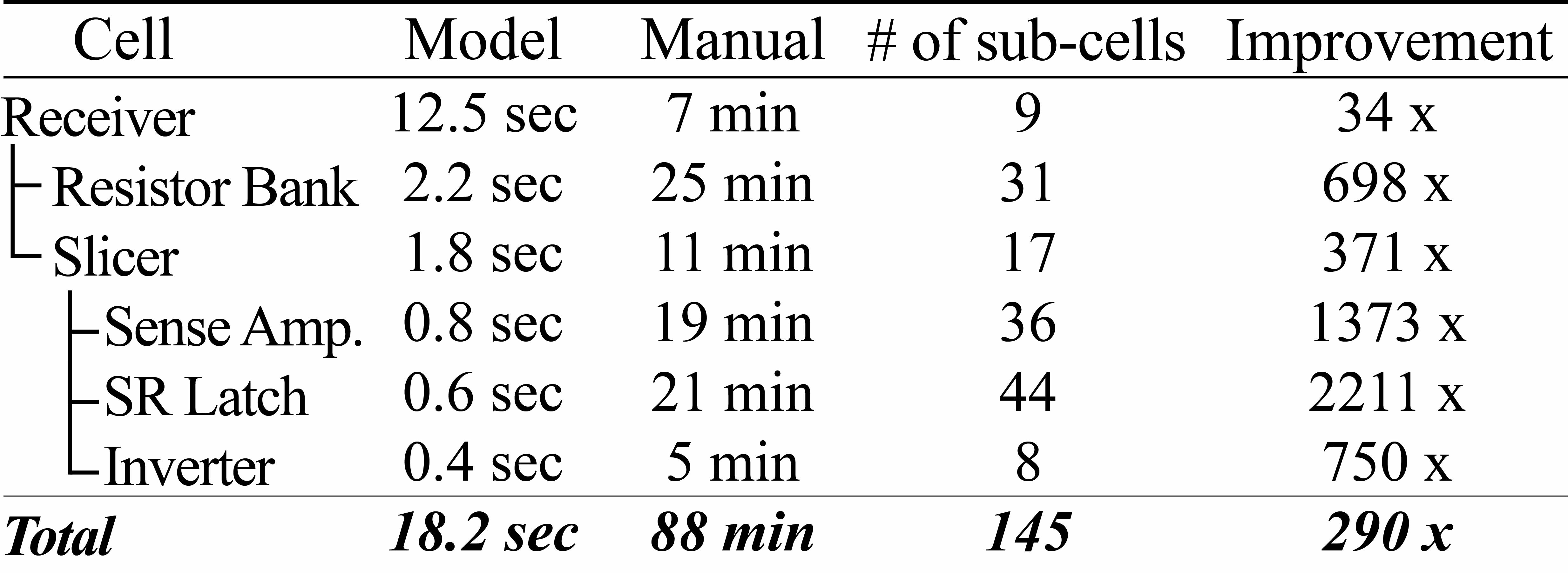}
\end{table}

In practical scenario, the layout-to-generator conversion follows the application’s construction cycle introduced in Section III. Sub-cell layouts are converted to sub-cell generators one by one, starting from the bottom of the design hierarchy. This process gradually fills the generator library with sub-cell generators, starting from small and frequently-used sub-cells in the lower design hierarchy. As new generators are added, the CNN model is also updated by training with new layouts generated by these new generators. As a result, the model becomes familiar with these new sub-cell types and can classify them better. Because the layout-to-generator conversion starts from bottom of the design hierarchy, usually beginning with small sub-cell designs that are frequently instantiated, there is a high chance that the model is familiar with such small sub-cell designs during inferencing. In addition, typically, the majority of sub-cells are small and simple, and thus can be easily classified. For example, transistors, resistors, inverters, capacitors, current mirrors, etc. are such frequently-used sub-cells. Because their count is very large and they can be easily classified, the overall classification accuracy would be very high in practical scenario.  In Table IV (TRX), we reported the accuracy of classifying sub-cell designs: 535 sub-cell designs out of 557 different designs were correctly classified, resulting in 96.1\%. If we re-calculate the accuracy of classifying sub-cell instances, then 17,120 sub-cell instances out of total 17,206 sub-cell instances were correctly classified, resulting in the accuracy of 99.5\%. Therefore, the model’s accuracy is expected to be higher in the practical scenario, where classification of sub-cell instances matters.

Additionally, we compared the CNN model’s accuracy with that of a Support Vector Machine (SVM) as a baseline. We applied similar preprocessing steps to both models. For the CNN, we used a 21-channel 256x256 matrix. For the SVM, which requires one-dimensional vectors, we extracted histograms along the x and y axes for each channel. This produced 42 histogram vectors of length 256 (21 channels × 2 directions), allowing the SVM to use the same dataset as the CNN.

In the TRX test, the SVM did not identify any of the 69 not-generatable sub-cells, resulting in a Not-Generatable Identification Rate (NGIR) of 0\%. In contrast, the CNN model achieved an NGIR of 87.0\% for top-3 predictions and 78.3\% for top-1 predictions. The SVM could not recognize new and unfamiliar not-generatable layouts, while the CNN model handled them well. The NGIR is important because incorrect assignment of generators to not-generatable sub-cells can cause design rule or LVS violations. In addition, the CNN model also exceeded the SVM’s overall accuracy, achieving 96.1\% (98.2\% in top-3) compared to 80.1\% for the SVM. These results show that the CNN model accurately classifies not-generatable layouts, even for sub-cells designed in different technology nodes or not included in the training dataset.

\subsection{Runtime Comparison with Manual Work}

\begin{table}[t]
        \caption{RUNTIME BREAKDOWN TABLE OF THE PROPOSED CNN MODEL.}
        \label{tab5}
        \centering
        \includegraphics[width=0.85\linewidth]{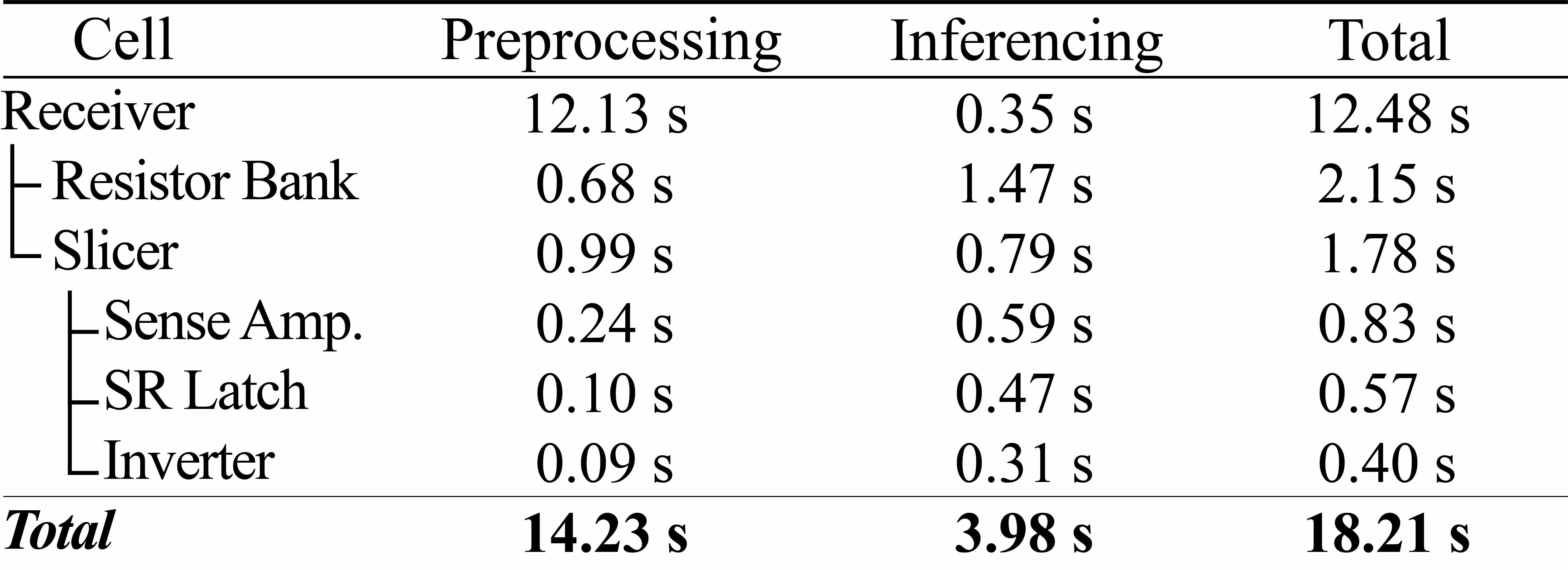}
\end{table}
        
\begin{table}[t]
        \caption{RUNTIME BREAKDOWN TABLE OF THE BASE MODEL.}
        \label{tab6}
        \centering
        \includegraphics[width=0.85\linewidth]{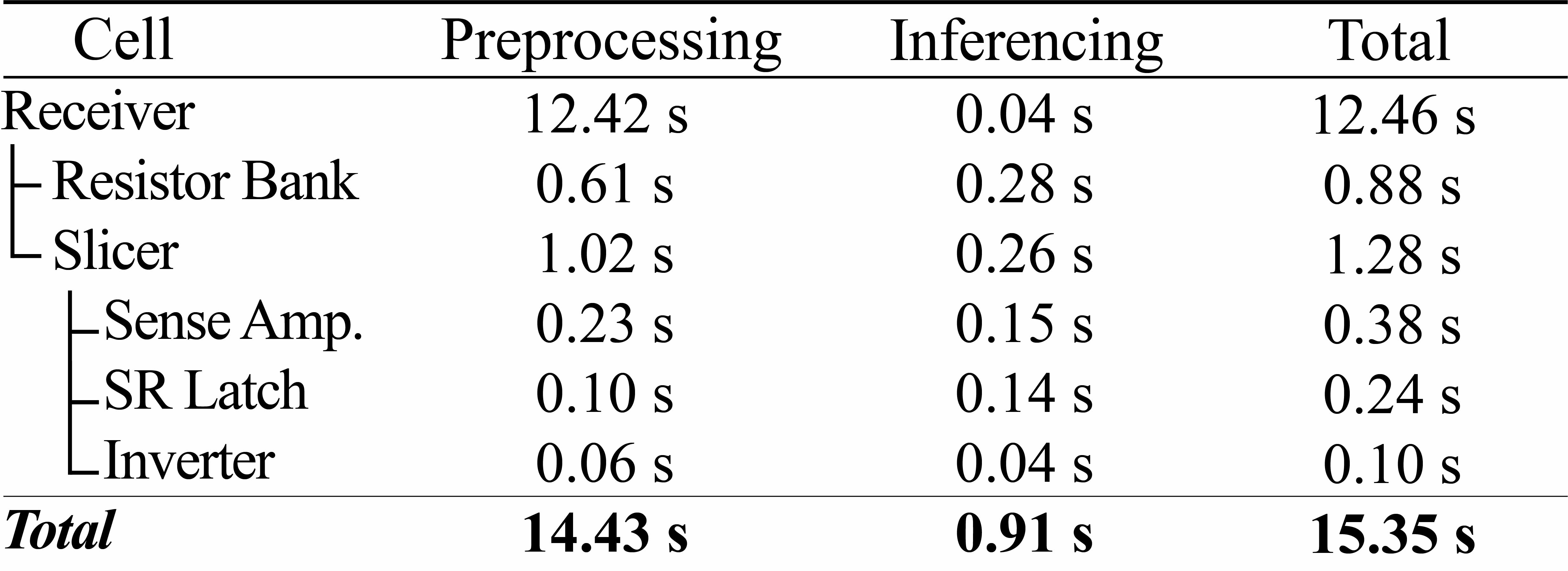}
\end{table}

We also measured and compared the time required for the proposed CNN model and a human designer to classify sub-cells of the RX. In this experiment, we assumed the layout-to-generator conversion scenario explained previously in Section III.

For the same layout design, the CNN model and a human designer did the same task: examining the generator assignment for the sub-cells. The model was implemented in the GUI-based layout-to-generator framework \cite{SUN} and ran on a personal computer with an Apple M1 Max chip, a 64 GB RAM rather than on a powerful GPU server.

To reduce unnecessary runtime, our algorithm avoids multiple examinations of the identical layout instances referencing the same design source by checking examination history. Therefore, the runtime for examination of sub-cell instances could be greatly reduced by examining only different sub-cell designs instead of examining all sub-cell instances.

Table V shows the measured times for examining the sub-cells of the RX circuit. The result shows that examination time was greatly reduced by the CNN model. The model only took 18.2 seconds for preprocessing, inferencing, and suggesting generator scripts for 4,885 sub-cell instances. Because 4,885 sub-cell layout instances are referencing 145 sub-cell design sources, all 4,885 sub-cells of RX were classified by examining only 145 sources within 18.2 seconds. If we examined all 4,885 sub-cell layout instances individually, it would take 69 seconds, taking approximately 3.8 times longer than examining the 145 sub-cell design sources. 

In contrast, an expert human designer spent 88 minutes manually examining 145 sub-cells. This result shows that the CNN model could reduce the time required for suggesting the correct generator scripts by 290 times. Examining hundreds of sub-cells and tens of generators makes a human designer easily exhausted because, for this task, the designer must compare the layout patterns of the sub-cells and generators in detail. It is noticeable that the experimental setup was very favorable to the human designer. He was familiar with both programming and the RX circuit, and had prior knowledge of which sub-cells could be generated by which available generators. A practical scenario described in Section III would be much worse for the human designer. In this test, the number of classes of generation possibilities was only 52. However, in a practical scenario described in Section III, there would be a lot more available generators. Therefore, this task would be much more time-consuming. If the designer has many more layout generator scripts to choose from and has less prior knowledge about the sub-cell layout, then examining sub-cell layouts and finding the matched generator scripts would be much more time-consuming for the human designer.

Table VI shows the detailed time breakdown. Larger layouts take a longer time to be converted into a matrix, but the process generally completes in tens of seconds. This is acceptable given that the application is focused on small sub-cells. Authors believe that in the future, if the application requires the handling of larger cells, the time of the matrix transformation process can be reduced by parallel algorithms.

\subsection{Impact of Model Architecture}

\begin{figure}[t]
\centering
\includegraphics[width=0.9\linewidth]{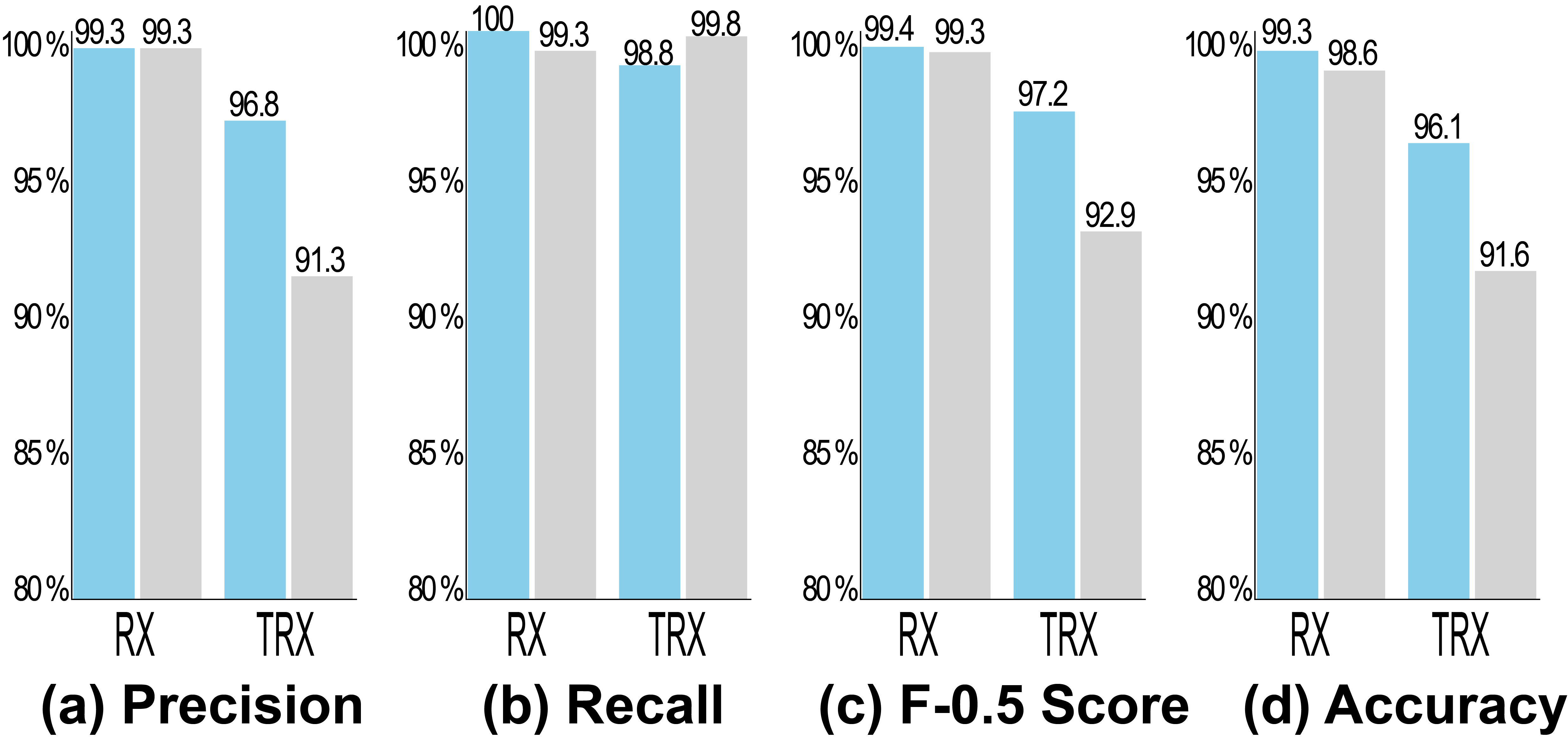}
\caption{Performance comparison between the proposed CNN model and the base CNN model.}
\label{fig12_comparison}
\end{figure}

We compared the performances and runtimes of two CNN architectures: the proposed CNN model with sub-networks and the base CNN model.

Fig. 13 shows the precision, recall, F-0.5 score, and accuracy results of the two model architectures for two different datasets. For RX, the proposed CNN model achieved an F-0.5 score of 99.4\%, while the base CNN model achieved 99.3\%. This high performance indicates that the CNN architectures are good at identifying sub-cell patterns when the test dataset is similar to the training dataset. When examining unfamiliar layouts of TRX that are not similar with the training dataset, the proposed CNN model with three sub-networks outperformed the base CNN model. Specifically, the proposed model achieved a F-0.5 score of 97.2\%, compared to the base model’s score of 92.9\%. 

The runtime of the proposed model is slightly longer than that of the base model. To classify RX's sub-cells, the proposed model took 18.21 seconds, requiring only 2.86 seconds more than the base model (Table VII). The overall runtime increase by the proposed CNN was contributed by the increase of the inferencing time. Because the proposed CNN mode is more complex than the base model, it requires more computation for inferencing, taking 3.07 second more. It is natural that data preprocessing times of both models are almost identical because they conduct the same data preprocessing. Although the inferencing time of the proposed CNN is much larger than the base model’s, the overall runtimes of both models are not substantially different because data preprocessing times are dominant in the total computation times of both models. 

\begin{figure}[b]
        \centering
        \includegraphics[width=\linewidth]{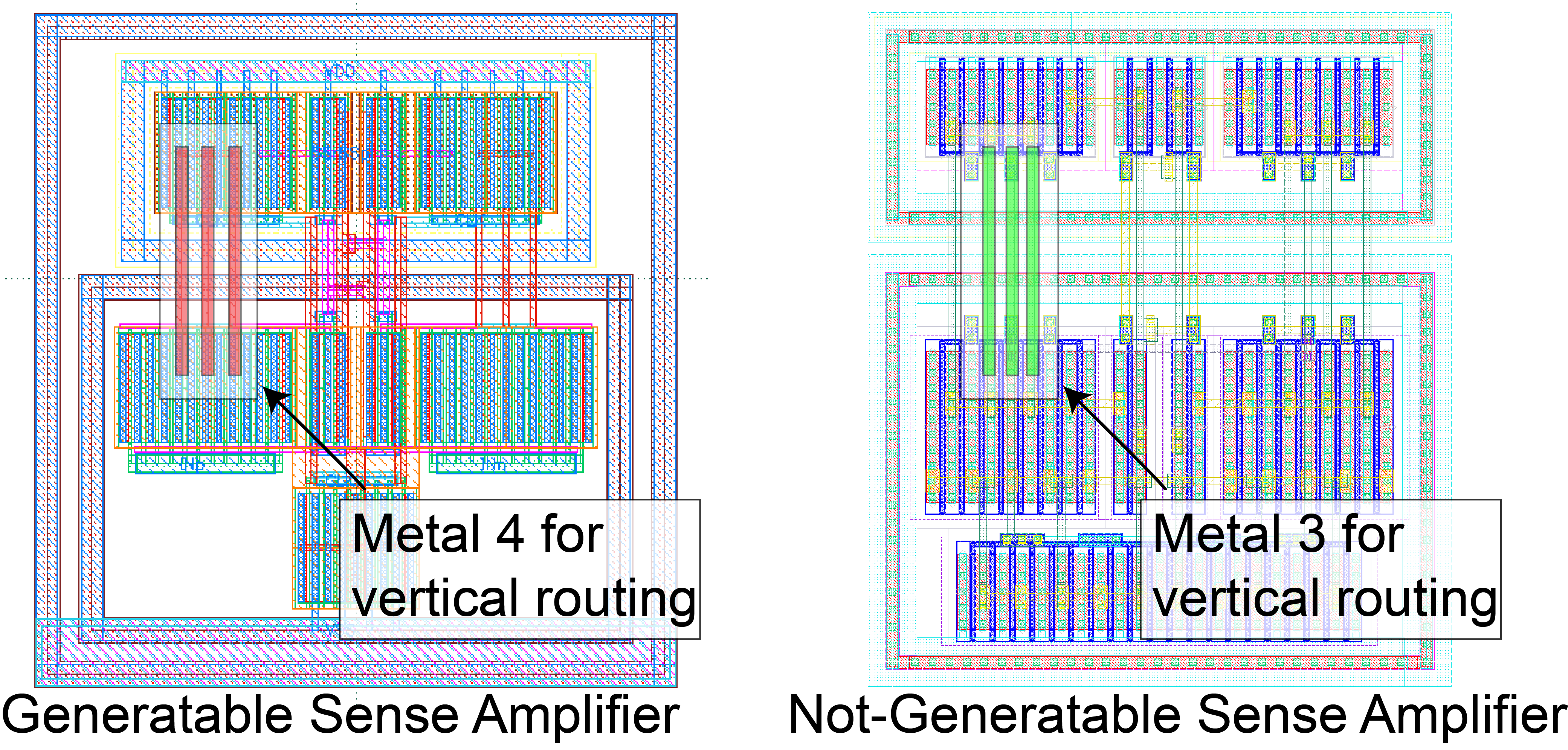}
        \caption{Generatable and not-generatable sense amplifiers distinguished by the model. They look similar in a placement style, but their routing styles (metal direction) are different.}
        \label{fig13_tricky_sa}
\end{figure}

\begin{figure}[t]
        \centering
        \includegraphics[width=\linewidth]{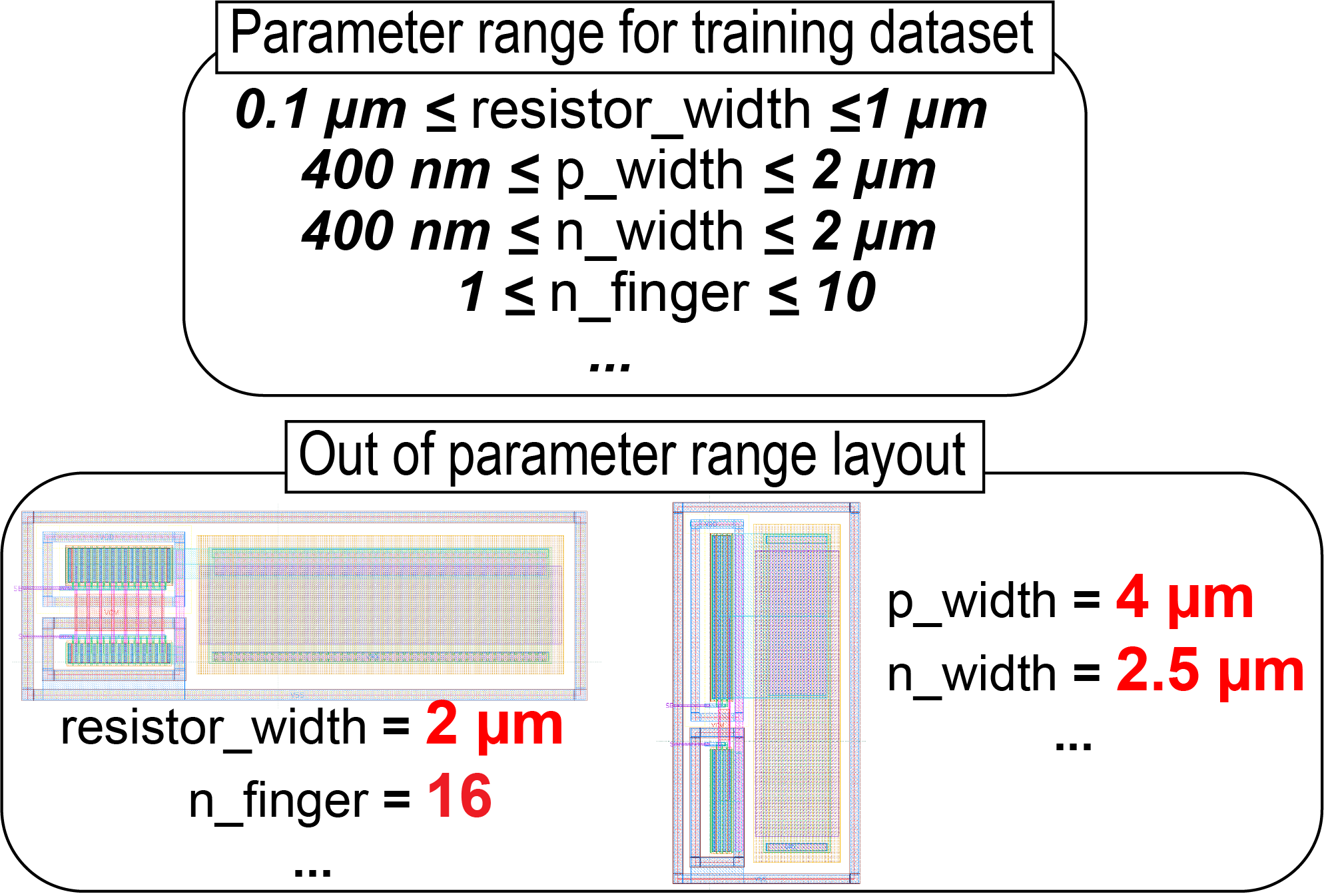}
        \caption{Examples of correctly inferred two resistor bank units that are not within the design space of the trained dataset. They were generated by the same generator using different design parameters. They were correctly classified by the model even though their design parameters are not within the ranges of design parameters of the training dataset.}
        \label{fig14_tricky_parm}
\end{figure}

\subsection{Classification of Tricky Layouts}
It is interesting that the model could solve tricky layout classification problems that can easily confuse a human designer. Fig. 14 shows two sense amplifiers in different layout styles. The first one can be generated by a sense amplifier generator used in this work whereas the second one cannot. Both layouts look similar to human designers because of the similarity in placement. However, their metal routing directions (vertical or horizontal) are different. The model could distinguish the subtle differences in metal routing and classify them correctly. Another interesting observation is that the model could correctly recognize them if two layouts are generated by the same generator even though they have very different aspect ratios and design parameters. Fig. 15 shows two resistor bank units that were generated by the same generator with different design parameters. They look very different to human designers, but the model could recognize them as the same class. Furthermore, the model could correctly recognize layouts that are not within the design space of the training dataset. The design parameters of the two resistor bank units in Fig. 15 are not within the ranges of design parameters of the training dataset: the model has not been trained with layouts with such an extreme aspect ratio. Even though, the model could manage to recognize that they were generated by a generator of a resistor bank unit.

\subsection{Comparison with Other Works}
        
\begin{table*}[t]
        \caption{COMPARISON TABLE WITH OTHER WORKS.}
        \label{tab6}
        \centering
        \includegraphics[width=\linewidth]{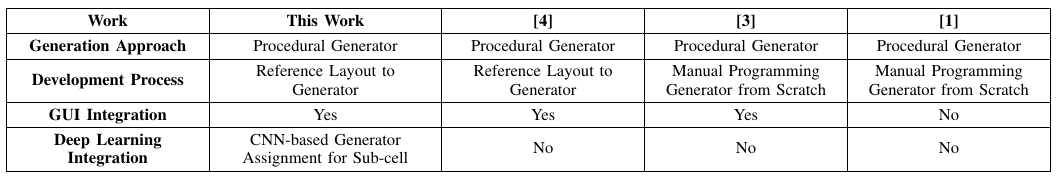}
\end{table*}

To the best of our knowledge, our sub-cell layout classification and generator assignment method has not been previously studied, making direct comparisons with prior works challenging. Nevertheless, we compare our proposed method with existing layout generation frameworks in Table VIII.

Jeong et al. [4] provides a layout-to-generator conversion process but relies on manual generator assignment for sub-cells, which is a time-consuming and error-prone process. Han [3] offers a GUI-based development environment but cannot convert reference layouts into procedural layout generators. Chang [1] proposed an open-source procedural generator framework that supports procedural generator development through manual programming. However, it does not provide layout-to-generator conversion or a GUI.

\section{Conclusion}
We automated the assignment of generator scripts to sub-cells to assist layout-to-generator conversion. To examine whether sub-cells can be generated by available generator scripts or not, we propose a CNN model that can examine various sub-cell layouts. Our proposed preprocessing method converts a layout input into a matrix of 256x256x21 after layer mapping. The proposed CNN model employs three parallel sub-networks with input sizes of 64x64, 128x128, and 256x256 in order to extract and process multi-scale feature maps. By this multi-scale approach, the proposed model can achieve good accuracy for various sizes of input sub-cells. 

In experiment, the CNN model was trained with a dataset generated by available generators. Tens of thousands of layout data were automatically generated with random design parameters for training. In examining the 145 different sub-cell designs among a total 4,885 sub-cell instances of a high-speed wireline receiver, the CNN model inferred the generator suggestion correctly for 144 sub-cell designs and incorrectly for only one sub-cell design, achieving a F-0.5 score of 99.4\%. The time for examining sub-cells was greatly reduced to 18 seconds by the CNN model from 88 minutes by manual checking: this corresponds to 290 times improvement in examination efficiency. These results show that artificial intelligence potentially can greatly help layout-to-generator conversion. The CNN model assists converting layouts to the generators which can generate many new layouts and strengthen the CNN model. With this positive feedback, the layout and the layout generator libraries can be easily and quickly expanded in the future.

\section*{Acknowledgments}
The EDA tool was supported by the IC Design Education Center(IDEC), Korea.


\bibliographystyle{IEEEtran}

\bibliography{IEEEabrv,reference.bib}

% \bibitem{ref1}
% {\it{Mathematics Into Type}}. American Mathematical Society. [Online]. Available: https://www.ams.org/arc/styleguide/mit-2.pdf

% \bibitem{ref2}
% T. W. Chaundy, P. R. Barrett and C. Batey, {\it{The Printing of Mathematics}}. London, U.K., Oxford Univ. Press, 1954.

% \bibitem{ref3}
% F. Mittelbach and M. Goossens, {\it{The \LaTeX Companion}}, 2nd ed. Boston, MA, USA: Pearson, 2004.

% \bibitem{ref4}
% G. Gr\"atzer, {\it{More Math Into LaTeX}}, New York, NY, USA: Springer, 2007.

% \bibitem{ref5}M. Letourneau and J. W. Sharp, {\it{AMS-StyleGuide-online.pdf,}} American Mathematical Society, Providence, RI, USA, [Online]. Available: http://www.ams.org/arc/styleguide/index.html

% \bibitem{ref6}
% H. Sira-Ramirez, ``On the sliding mode control of nonlinear systems,'' \textit{Syst. Control Lett.}, vol. 19, pp. 303--312, 1992.

% \bibitem{ref7}
% A. Levant, ``Exact differentiation of signals with unbounded higher derivatives,''  in \textit{Proc. 45th IEEE Conf. Decis.
% Control}, San Diego, CA, USA, 2006, pp. 5585--5590. DOI: 10.1109/CDC.2006.377165.

% \bibitem{ref8}
% M. Fliess, C. Join, and H. Sira-Ramirez, ``Non-linear estimation is easy,'' \textit{Int. J. Model., Ident. Control}, vol. 4, no. 1, pp. 12--27, 2008.

% \bibitem{ref9}
% R. Ortega, A. Astolfi, G. Bastin, and H. Rodriguez, ``Stabilization of food-chain systems using a port-controlled Hamiltonian description,'' in \textit{Proc. Amer. Control Conf.}, Chicago, IL, USA,
% 2000, pp. 2245--2249.

% \end{thebibliography}

\newpage

\vspace{11pt}

\begin{IEEEbiography}[{\includegraphics[width=1in,height=1.25in,clip,keepaspectratio]{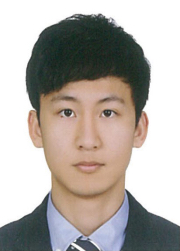}}]{Sungyu Jeong}
(Graduate Student Member, IEEE) received the B.S degree in electrical engineering from Pohang University of Science and Technology (POSTECH), Pohang, Korea, in 2018, where he is currently pursuing the Ph.D. degree. 

His current research interests include the layout automation of analog circuits, as well as design automation enhanced by deep learning.
\end{IEEEbiography}
% \vspace{-3cm}
\vspace{11pt}

\begin{IEEEbiography}[{\includegraphics[width=1in,height=1.25in,clip,keepaspectratio]{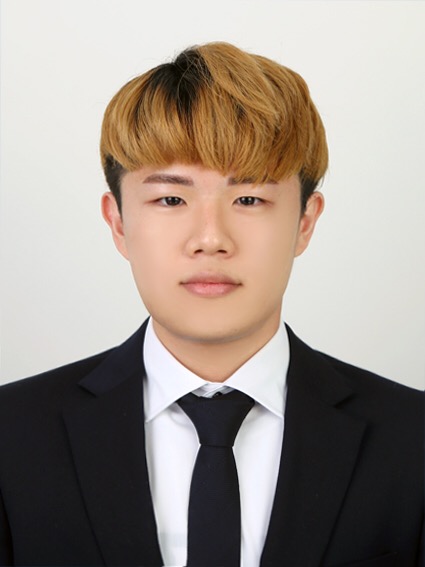}}]{Minsu Kim}
(Graduate Student Member, IEEE) received the B.S. degree in electronic and electrical engineering from Sungkyunkwan University (SKKU) ,Suwon, Korea, in 2020 and received the M.S. degree in the electronic and eleectrical engineering from Pohang University of Science and Technology (POSTECH), Pohang, Korea, in 2022, where he is currently pursuing the Ph.D. degree. His research interests include high-speed link circuit and computer aided design.
\end{IEEEbiography}
% \vspace{-3cm}
\vspace{11pt}

\begin{IEEEbiography}[{\includegraphics[width=1in,height=1.25in,clip,keepaspectratio]{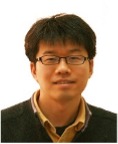}}]{Byungsub Kim}
(Senior Member, IEEE) received the B.S. degree in electrical engineering from the Pohang University of Science and Technology (POSTECH), Pohang, South Korea, in 2000, and the M.S. and Ph.D. degrees in electrical engineering and computer science from Massachusetts Institute of Technology (MIT), Cambridge, MA, USA, in 2004 and 2010, respectively.
He was an Analog Design Engineer with Intel Corporation, Hillsboro, OR, USA, from 2010 to 2011. In 2012, he joined the faculty of Department of Electrical Engineering, POSTECH, where he is currently a Professor.
Dr. Kim received several honorable awards. He received the IEEE Journal of Solid-State Circuits Best Paper Award in 2009. In 2009, he was an also co-recipient of the Beatrice Winner Award for Editorial Excellence at the 2009 IEEE International Solid-State Circuits Conference. For several years, he served as a member of Technical Program Committee of the IEEE International Solid-State Circuits Conference and has been serving as the Chair of Wireline Sub-com of the IEEE Asian Solid-State Circuit Conference.
\end{IEEEbiography}

\vfill

\end{document}